%% file: Jelly_orbits_v3.tex
\documentclass[useAMS,usenatbib,usedcolumn]{mn2e}
\input{mycommands}

\usepackage[figuresright]{rotating}
\usepackage{lscape}
\usepackage{graphics}
\usepackage{epsfig}
\usepackage{multirow}
\usepackage{bigdelim}
\usepackage{bigstrut}
\usepackage{amsmath}
\usepackage{amssymb}
\usepackage{hyperref}
\usepackage{epstopdf}
\usepackage{enumerate}
\usepackage{color}

\defcitealias{Jaffe2016}{J16}

\title[GASP IX. Jellyfish galaxies in phase-space]{GASP IX. Jellyfish galaxies in phase-space: an orbital study  of intense ram-pressure stripping in clusters\\
}
\author[Y.~Jaff\'e et al.] {Yara L. Jaff\'e$^{1}$\thanks{E-mail: yjaffe@eso.org}, 
Bianca M. Poggianti$^{2}$,
Alessia Moretti$^{2}$, 
Marco Gullieuszik$^{2}$,  \and
Rory Smith$^{3}$, 
Benedetta Vulcani$^{4,2}$,
Giovanni Fasano$^{2}$, 
Jacopo Fritz$^{5}$, \and
Stephanie Tonnesen$^{6}$, 
Daniela Bettoni$^{2}$, 
George Hau$^{1}$, 
Andrea Biviano$^{7}$, \and
Callum Bellhouse$^{8}$,
Sean McGee$^{8}$\\
   $^1$European Southern Observatory, Alonso de Cordova 3107, Vitacura, Casilla 19001, Santiago de Chile, Chile \\
   $^2$INAF - Osservatorio Astronomico di Brera, via Brera 28, 20122 Milano, Italy\\
   $^3$ Korea Astronomy and Space Science Institute, 766, Daedeokdae-ro, Yuseon-gu, Daejon, 34055, Korea \\
   $^4$ School of Physics, The University of Melbourne, VIC 3010, Australia.  \\
   $^5$ Instituto de Radioastronomía y Astrof\'isica, Morelia, Mexico \\
   $^6$ Center for Computational Astrophysics, Flatiron Institute, 162 5th Ave, New York, NY 10010, USA \\
   $^7$ INAF - Astronomical Observatory of Trieste, 34143 Trieste, Italy\\
   $^8$ University of Birmingham School of Physics and Astronomy, Edgbaston, Birmingham, England\\
   }
\begin{document}

\maketitle

\begin{abstract} 
It is well known that galaxies falling into clusters can experience gas stripping due to ram-pressure by the intra-cluster medium (ICM). The most spectacular examples are galaxies with extended tails of optically-bright stripped material known as ``jellyfish''.  
We use the first large homogeneous compilation of jellyfish galaxies in clusters from the WINGS and OmegaWINGS surveys, and follow-up MUSE observations from the GASP MUSE programme to investigate the orbital histories of jellyfish galaxies in clusters and reconstruct their stripping history through position vs. velocity phase-space diagrams.
We construct analytic models to define the regions in phase-space  where ram-pressure stripping is at play.  
We then study the distribution of cluster galaxies in phase-space and find that jellyfish galaxies have on average higher peculiar velocities (and higher cluster velocity dispersion) 
than the overall population of cluster galaxies at all clustercentric radii, which is indicative of recent infall into the cluster and radial orbits.  In particular, the jellyfish galaxies with the longest gas tails reside very near the cluster cores (in projection) and are moving at very high speeds, which coincides with the conditions of the most intense ram-pressure. 
We conclude that many of the jellyfish galaxies seen in clusters likely formed via fast ($\sim 1-2$ Gyr), incremental, outside-in ram-pressure stripping during first infall into the cluster in highly radial orbits.
\end{abstract}


\section{Introduction}
\label{sec:introduction}

Numerous studies have shown that cosmic environment plays a key role transforming galaxies and quenching their star formation  \citep[see][and references therein]{BoselliGavazzi06}. The physical mechanisms responsible for such transformations can be divided into two main types: 

(i) Gravitational or tidal mechanisms can affect both the stars and the gas within galaxies. They include: galaxy-galaxy interactions and mergers \citep{BarnesHernquist1996}, interactions between galaxies and the gravitational potential of the host cluster \citep{Byrd1990}, and the so-called ``harassment'', or cumulative high-speed encounters between galaxies \citep{Moore1996}

(ii) Hydrodynamical interactions between cluster galaxies' gaseous component and the hot intra-cluster medium (ICM) can result in gas stripping due to ram pressure \citep{GunnGott1972}, thermal evaporation \citep{CowieSongaila1977}, or turbulent/viscous stripping \citep{Nulsen1982}.

There is mounting evidence that ram-pressure stripping is one of the most effective mechanisms quenching galaxies in clusters. Simulations show that ram-pressure stripping can effectively remove cold gas from galaxies, and in some cases temporarily enhance the star-formation activity before quenching it completely \citep[see e.g.][and references therein]{Steinhauser2016}. The most striking evidence supporting ram-pressure stripping comes from observations of neutral gas (HI) in nearby galaxies located inside clusters \citep{Haynes1984,Cayatte1990,Kenney2004,Chung2009,Vollmer2001,Jaffe2015,Yoon2017} or even groups \citep{Rasmussen2006,Rasmussen2008,VerdesMontenegro2001,Jaffe2012,Hess2013}. Such observations show how cold gas can indeed be removed from infalling cluster galaxies. 

In some cases, stars are formed in the stripped gas \citep{KenneyKoopman1999,Yoshida2008,Kenney2014}, and can thus be identified from UV or optical images. The most striking examples of stripped galaxies with new stars tracing the stripped tails are the so-called ``Jellyfish'' galaxies. 
Estimates based on small blind $H_{\alpha}$ surveys of Coma
and A1367 indicate that the fraction of cluster late-type galaxies with these features is close to 40\% \citep{BoselliGavazzi2014}.

Numerous studies have focused on individual jellyfish galaxies  \citep{Fumagalli2014, Ebeling2014, Rawle2014,Merluzzi2013,Fumagalli2014,Fossati2016}, and more recently, the first systematic searches for such objects at low and intermediate redshift have been conducted \citep{Poggianti2016,McPartland2016}. In particular, a dedicated ongoing integral-field spectroscopy survey with MUSE on the VLT has started to reveal with great detail the physics of gas removal processes in galaxies across a wide range of mass and environment \citep[GASP;][]{Poggianti2017a}. The first results of GASP confirmed the stripped nature of the jellyfish candidates observed so far, showing spectacular evidence for ram-pressure stripping in action \citep{Poggianti2017a,Bellhouse2017,Fritz2017,Poggianti2017b,Gullieuszik2017}. 

In addition to detailed observations and simulations of individual galaxies in clusters, a useful approach to study galaxy evolution in dense environments is to understand the assembly history of clusters. 
Pioneering work studying the position and velocities of different galaxy populations in clusters established the infall of spirals into clusters  \citep{CollessDunn1996,Mohr1996,Biviano1997}. A particularly useful tool to visualize the migration of galaxies from the field into clusters are thus position vs. velocity phase-space diagrams, as they reflect the virialization state of a cluster, and contain information about the orbital history of cluster galaxies. In the last decades, several works have studied the location of different populations of  cluster galaxies in projected phase-space to infer their time since infall  \citep{Vollmer2001, Mahajan2011, Oman2013,  Oman2016, HernandezFernandez2014, Haines2015, Muzzin2014, Jaffe2015, Jaffe2016, Yoon2017, Rhee2017}.

Despite known projection effects \citep[see discussion in][]{Oman2013,Rhee2017}, phase-space diagrams can further be used to identify different cluster mechanisms that affect galaxies during their passage, such as ram-pressure stripping of gas \citep[][]{Vollmer2001,Jaffe2015,Yoon2017} and tidal mass loss \citep{Rhee2017}. 
In particular \citet{Jaffe2015} compared the position of observed cluster galaxies with HI data with mock data to reconstruct the effect of ram-pressure stripping as a function of time since infall into the cluster. They show that significant gas stripping occurs on first infall and that galaxies gradually become HI-poor. In a follow up study \citet{Jaffe2016} further demonstrated that the gas removal from galaxies can be accompanied by a an initial enhancement of the star-formation activity, followed by a slow quenching \citep[see also][]{Haines2015}.

In this paper we investigate the orbital histories of jellyfish galaxies in low redshift clusters and the effect of ram-pressure stripping by the ICM. To achieve this, we compare the projected position and velocity of a large sample of cluster jellyfish candidates from \citet[][P16 from now onwards]{Poggianti2016} and a sub-sample of confirmed jellyfish galaxies from the GASP survey, with the control parent sample of cluster galaxies from WINGS \citep[WIde-field Nearby Galaxy-cluster Survey;][]{Fasano2006, Moretti2014} and its extension OmegaWINGS \citep{Gullieusik2015,Moretti2017}. 
In Section \ref{sec:data} we summarize the data available, and describe the sample of clusters, cluster galaxies, jellyfish candidates and confirmed jellyfish.  We also separate the jellyfish galaxies observed with MUSE into four different categories related to their stripping stages.
In Section~\ref{sec:rps_model} we construct analytic models of ram-pressure stripping for two (low and high-mass) clusters and two (low and high-mass) galaxies, and identify the regions in position vs. velocity phase-space diagrams where ram-pressure stripping is effective at removing gas from the galaxies in the different cases. 
In Section~\ref{sec:orbits} we use the constructed models to interpret the spatial and velocity distribution of jellyfish galaxies in relation to their stripping stage. Moreover, we compare the velocity distributions of different galaxy populations to study their virialization states and their orbital types. 
In Section~\ref{sec:conclu} we summarize our findings and draw conclusions. 

Throughout this paper we assume a concordance $\Lambda$CDM cosmology with $\Omega_{\rm M}=$0.3, $\Omega_{\Lambda}=$0.7, and H$_{0}=$70 km s$^{-1}$ Mpc$^{-1}$. We adopt a \citep{Chabrier2003} initial mass function (IMF) in the mass range 0.1-100$M_{\odot}$.


\begin{figure*}
\centering

\includegraphics[width=0.48\textwidth]{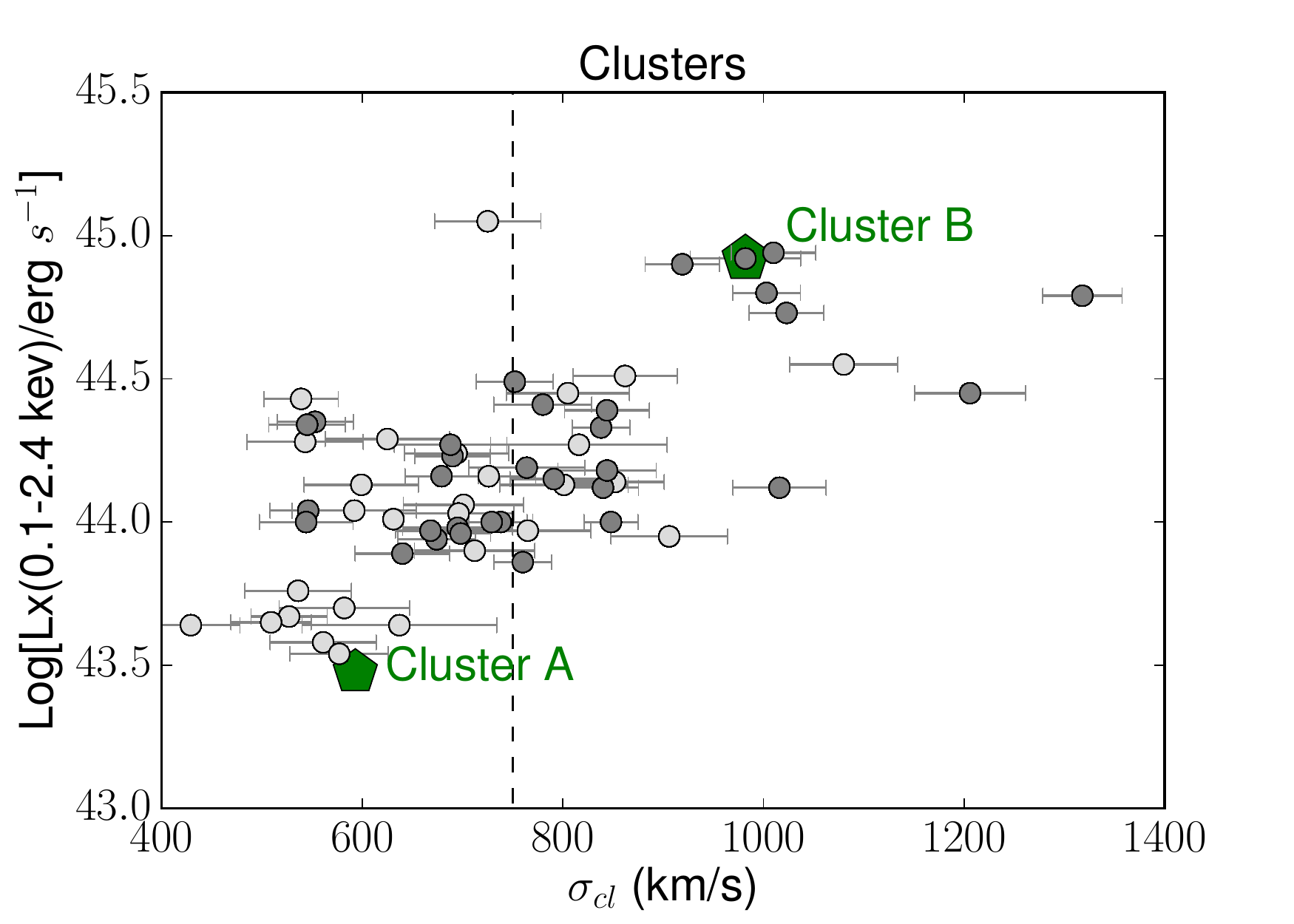}
\includegraphics[width=0.48\textwidth]{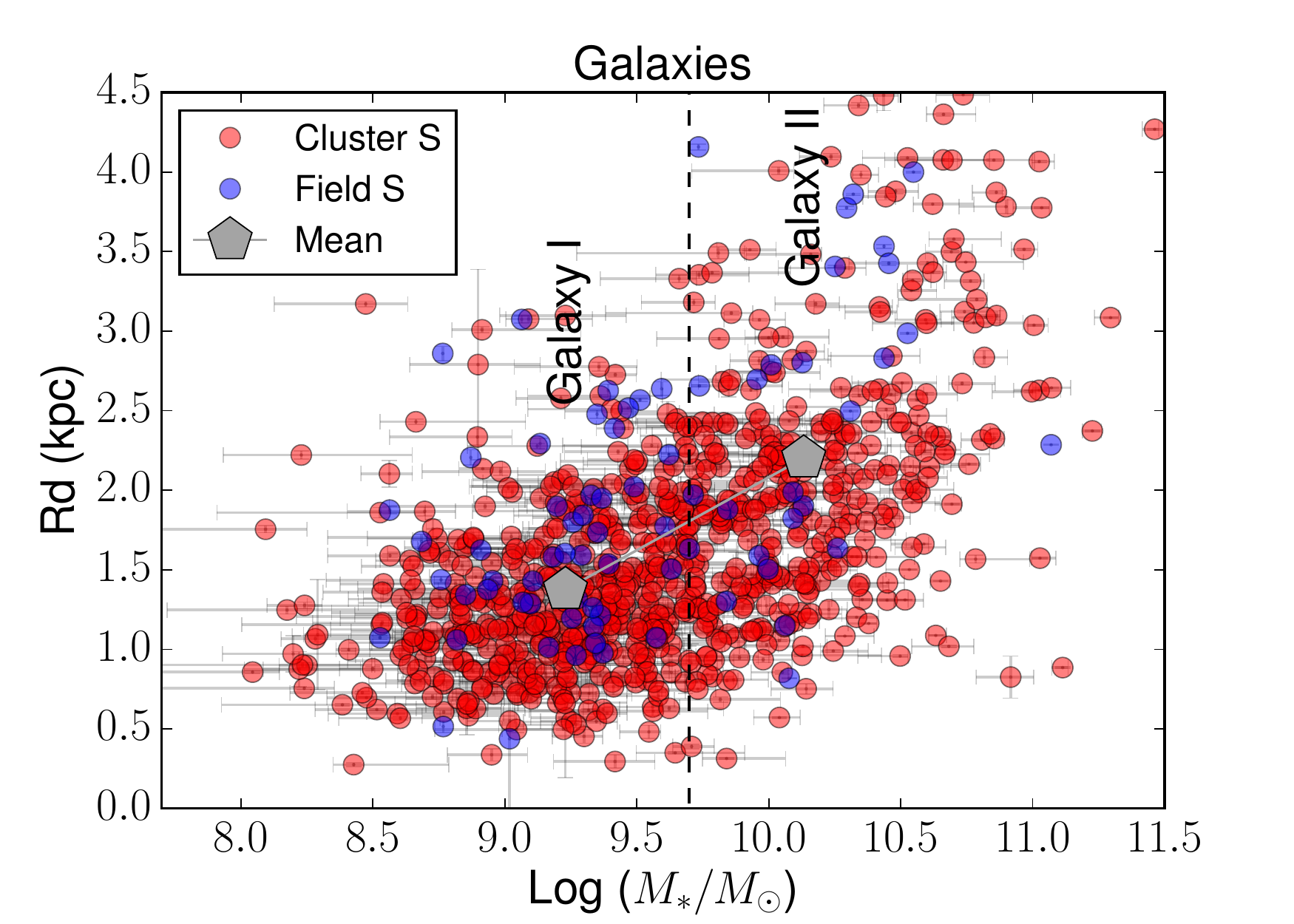}
\caption{ Left: cluster velocity dispersion ($\sigma_{cl}$, with error bars) vs. the X-ray luminosity ($L_{X}$)  of the WINGS/OmegaWINGS clusters used in this work (all grey points). Clusters with spectroscopic completeness $>50\%$ are highlighted with darker grey symbols. Right: The stellar mass ($M_{\bigstar}$) vs disk scale length ($R_d$) of all the cluster and field spirals (S) from the  WINGS/OmegaWINGS samples with a redshift. Error bars for both quantities are plotted, but note that for $R_d$ the errors are almost always smaller than the symbol.  The green pentagons indicate the properties of the model clusters A and B (left) and model galaxies I and II (right) discussed in Section~\ref{sec:Pram} (see also Tables~\ref{cluster_models} and~\ref{galaxy_models}). \label{cl_properties} }
\end{figure*}

\section{dataset}
\label{sec:data}

\subsection{The galaxy clusters}

This paper makes use of the sample of cluster galaxies from  
WINGS \citep{Fasano2006,Moretti2014} and its extension OmegaWINGS \citep{Gullieusik2015,Moretti2017}.

WINGS is a multi-wavelength survey of 76 galaxy clusters at $0.04<z<0.07$, selected in X-rays  from the ROSAT All Sky Survey  \citep{Ebeling1996,Ebeling1998,Ebeling2000}. 
The core of the survey are B and V imaging \citep{Varela2009}  that cover the central $34^{\prime}\times34^{\prime}$ of the clusters, reaching out to at least 0.6$\times R_{200}$\footnote{R$_{200}$ is defined to be the radius delimiting a sphere with interior mean density 200 times the critical density. It is typically used as an approximation for the cluster virial radius}. Complementary spectroscopic observations were obtained for a sub-sample of 48 clusters with WYFFOS@WHT and 2dF@AAT \citep{Cava2009}.

OmegaWINGS is an extension of WINGS that quadruples the area covered for a sub-sample of the clusters, allowing to study the effect of cluster environment on galaxies well beyond the virial radius.  Imaging in the u, B, and V bands were obtained with OmegaCAM@VST for 46 of the WINGS clusters over an area of $\sim$1~deg$^2$  \citep{Gullieusik2015}. 
The spectroscopic follow-up has been obtained for a sub-sample of 33 clusters with the 2dFdr@AAT \citep{Moretti2017}.

The clusters span a wide range of velocity dispersion \citep[$\sigma \sim$500-1300 km/s;][]{Cava2009,Moretti2017} and X-ray luminosity \citep[$L_{X}\sim 0.2-5 \times 10^{44}$ erg/s][]{Ebeling1996, Ebeling1998, Ebeling2000}, as shown in the left panel of Figure~\ref{cl_properties}. 
Since the clusters also span a range of  redshifts ($0.04<z<0.07$), and thus sizes in the sky, the radial coverage and completeness of the spectroscopy can vary significantly. However, the sample of all WINGS/OmegaWINGS clusters covers  at least $\sim 1\times R_{200}$ and up to $\sim 3.5 \times R_{200}$.  
32 out of the 76 clusters have a global spectroscopic completeness (calculated both as a function of V magnitude and radial projected distance from the BCG) higher than $\approx$ 50\% \citep[see Table 1 in][]{Paccagnella2017}. In this paper we use the full sample of WINGS/OmegaWINGS clusters (unless otherwise specified) noting that at clustercentric distances $> R_{200}$ biases originated by spectroscopic incompleteness could be present.

\subsection{The galaxy population}

In the fields of the WINGS/OmegaWINGS clusters there are 42,816 galaxies with a spectroscopic redshift. 17,957 of them are classified as cluster members, based on the $3 \sigma$ clipping method.

Morphologies of the WINGS galaxies were obtained applying the MORPHOT classification tool to the images,  that closely reproduces the visual classification of ellipticals, lenticulars and spirals \citep[][]{Fasano2012, Moretti2014}. 
Effective radii ($r_{eff}$) from single-Sersic fits were obtained using the GASPHOT tool  \citep{Pignatelli2006,DOnofrio2014}. Since this paper focuses on infalling disk galaxies (modelled in Section~\ref{sec:PI}), we compute the disk scale-length ($r_d$) for the disk galaxies (Sersic index $n\sim 1$) with spiral morphologies (T-type $>0$ from MORPHOT) assuming  $r_d = r_{eff} / 1.6783$, which is true for $n=1$.

Other galaxy properties were derived fitting the fiber spectra with the spectro-photometric model SINOPSIS \citep[SImulatiNg OPtical Spectra wIth Stellar population models;][]{Fritz2007, Fritz2011,Fritz2014}.  
The model utilizes the stellar population synthesis technique to reproduce the observed spectra. 
The output of the model includes star formation rates (SFRs) and stellar masses ($M_*$), among other quantities. 
In this paper we use stellar masses locked into stars that are in the nuclear-burning phase plus stellar remnants. 
The stellar masses of the cluster galaxies ranges between $10^{8}$ and $10^{12} M_{\odot}$ (mean $6.3 \times 10^{9} M_{\odot}$). The right panel of Figure~\ref{cl_properties} shows the galaxy mass as a function of size for the spiral galaxy population in and outside the clusters. 
The correlation between galaxy mass and size has a scatter between $\sim0.5$ and $0.9$~kpc in the stellar mass range from  $10^{9}$ to $ 10^{11}M_{\odot}$.

\begin{figure*}
\centering
\includegraphics[width=1\textwidth]{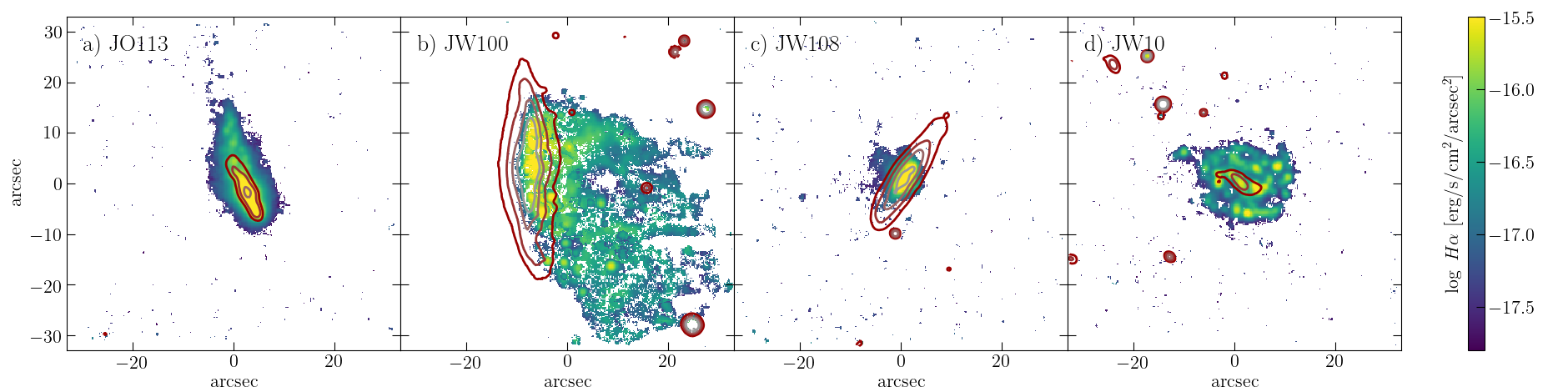}
\caption{H$\alpha$ flux emission maps (colourbar) on top of isophotes of  H$\alpha$ continuum in steps of 0.5 mag/arcsec$^{-2}$ (red contours) for 4 example GASP galaxies in different stripping stages. From left to right: a) A galaxy with moderate signs of  ``Stripping'';  b) a galaxy featuring very long gas tails indicative of ``Extreme-stripping''; c) a truncated ``Post-stripping'' galaxy; and d) a disturbed galaxy where the physical cause of the disturbance is unclear (``?''). 
\label{examples}}
\end{figure*}

\subsection{The sub-sample of jellyfish candidates}

The focus of this paper is to study the environmental history of jellyfish or heavily stripped galaxies in clusters. We use the jellyfish candidate catalog of P16, that contains 409 nearby galaxies with optical signatures suggestive of gas stripping in a wide range of environments from the samples of WINGS, OmegaWINGS, and PM2CG \citep{Calvi2011}, a field comparison sample, not considered in this paper. 
The jellyfish galaxy candidates were split into distinct ``classes'' (JClass) from visual inspection of B-band images: 
\begin{itemize}
 \item JClass 5 and 4 correspond to the most convincing cases of stripping. They include the  classical jellyfish galaxies with clear t
 entacles of stripped material.
 \item JClass 3 are probable cases of stripping with clear one-sided asymmetries.
 \item JClass 2 and 1 are the weakest cases  of stripping.
\end{itemize}

P16 showed that the main bodies of the jellyfish candidates are mostly disky, with stellar masses ranging from $10^9$ to $10^{11.5} M_{\odot}$. The same authors showed that, although the jellyfish galaxies span a similarly broad range of stellar masses and host cluster mass as the parent sample of WINGS/OmegaWINGS galaxies, they show enhanced star-formation rates (by a factor of $\gtrsim$ 2) compared to non-stripped galaxies of similar stellar mass. 
Of> the jellyfish candidates with a spectroscopic redshift, 70\% are cluster members. However, P16 found that the incidence of jellyfish with a spectroscopic redshift from WINGS/OmegaWINGS did not show a clear correlation with cluster velocity dispersion or X-ray luminosity.

\subsection{Confirmed jellyfish galaxies}

The ongoing GASP survey will observe over 100 jellyfish candidates from P16 as well as a control sample of non-stripped galaxies. So far, 80 of the GASP galaxies have been observed with MUSE.  Here we focus on the sub-sample of 49 of the GASP jellyfish in clusters observed to date.

From the MUSE datacubes, we have created maps of the stellar (continuum) and  gas (H$_{\alpha}$) components of the galaxies \citep[see examples in Fig~\ref{examples} and in][]{Poggianti2017a,Poggianti2017b,Bellhouse2017,Fritz2017,Gullieuszik2017}. These maps were used to  visually 
re-classify the galaxies  according to their apparent stripping stage into the following categories:
\begin{itemize}
 \item ``Extreme-stripping'': long one-sided gas tails comparable or larger than the stellar diameter.
 \item ``Post-stripping'': truncated gas disks that are smaller than the stellar disk. 
  \item ``Stripping'': any other clear sign of stripping, such as the presence of a disturbed gaseous component relative to the stars.
 \item ``?'': complex disturbed morphology that does not allow to conclude that ram-pressure stripping is the (only) mechanism at play. A couple of these cases show galaxies nearby that could potentially be interacting with them.
\end{itemize}
We emphasize that the component that is disturbed in the jellyfish galaxies is the gas. The stars are undisturbed and rotate smoothly within the disk \citep[not shown here, but see][]{Poggianti2017a,Bellhouse2017,Gullieuszik2017, Fritz2017}, which supports the idea that its ram-pressure stripping rather than tidal effects giving rise to the jellyfish morphology.   
Out of the 49 cluster galaxies observed by GASP, we found that 22 are experiencing moderate gas stripping (``Stripping''), 12  ``Extreme-stripping'' and 4 ``Post stripping''. The remaining 16 show complex morphologies (``?'').
Figure~\ref{examples} shows composite images from MUSE data of a few examples jellyfish galaxies observed by GASP in the different categories. The effect of the ram-pressure ``wind'' is visible from the $H_{\alpha}$ morphologies of the stripped galaxies in panels a, b, and c. Moreover, the fact that jellyfish galaxies still hold gas in their cores is indicative of outside-in stripping. This hypothesis is also supported by the gas kinematics and metallicity gradients seen in jellyfish galaxies \citep{Poggianti2017a,Bellhouse2017,Gullieuszik2017,Fritz2017}. 

From the classification exercise it became clear that
the distribution of H$\alpha$ gas with respect to the stellar component (this paper) is a much more representative and quantitative way to determine the stripping stage of a galaxy than through broad-band (BB) imaging (P16), which only shows the tip of the iceberg of the stripped gas. 
Nevertheless, the classification of GASP galaxies is mostly in agreement with the JClass classification of P16:  GASP galaxies visibly undergoing ``Extreme stripping''  also have the highest JClass numbers, while the ``stripped'' and ``post-stripped'' galaxies have JClass$\sim 3$. 
However, weak jellyfish features in BB optical images are not able to distinguish between stripping and post-stripping.

In this work we treat the confirmed jellyfish galaxy sample (i.e. galaxies observed by GASP so far that show signs of stripping) separately from the bigger sample of jellyfish candidates from \citet{Poggianti2016}, and utilize MUSE data when available to compute spectroscopic redshifts (and thus cluster membership), and stellar masses. The redshift was obtained from the spectra in the central part of the galaxy, while the masses were computed from SINOPSIS, using the integrated spectra within the lowest surface brightness stellar isophote defined from the continuum underlying $H_{\alpha}$, down to $\sim 1 \times \sigma$ from the background level \citep[see sec.6.5 in][for details]{Poggianti2017a}.

Finally, it is important to note that the detectability of stripped gas tails (and thus jellyfish galaxies) is expected to be sensitive to the lifetime of the tail before it is mixed (which is very poorly understood), as well as the inclination of the galaxy motion with respect to the observer. Between galaxies being stripped along the line of sight and galaxies moving along the plane of the sky, the latter are more likely to display extended visible tails that could facilitate their identification as ``jellyfish''. However, we note that we have confirmed cases of line-of-sight stripping in our sample. The most extreme of such examples is the GASP galaxy JO201, presented in \citet{Bellhouse2017}, which not only has clear tails of stripped gas along the plane of the sky (due to a small inclination angle with respect to the observer), but also a complex gas kinematics that impressively shows the stripped material that is dragging behind the galaxy. However, most of the other galaxies undergoing ``Extreme stripping'' \citep[e.g. JO204:][]{Gullieuszik2017}, are more likely to be traveling mostly in a direction perpendicular to the observer.

\begin{figure}
\centering
\includegraphics[width=0.49\textwidth]{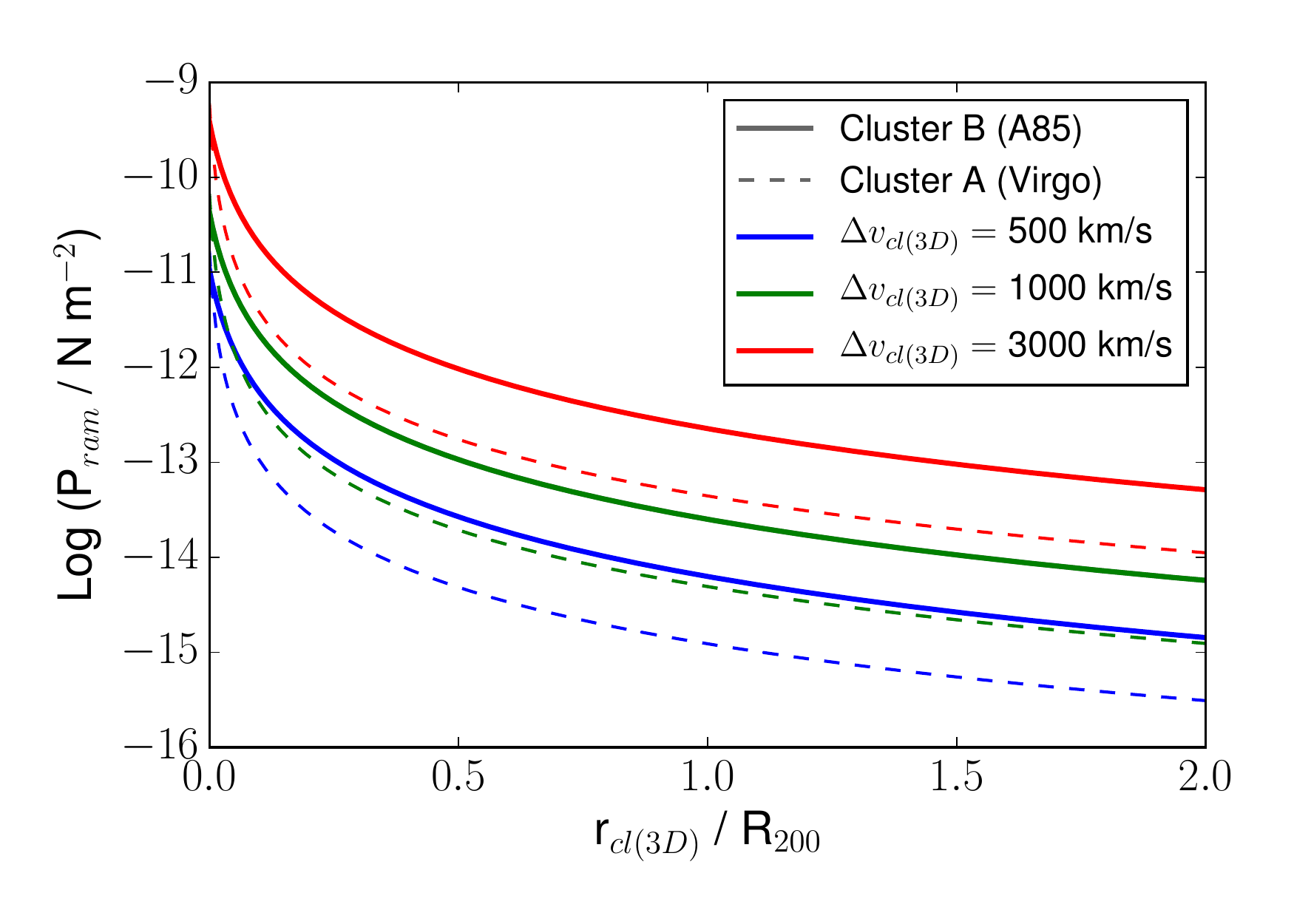}
\caption{The ram-pressure intensity profile for the model clusters A (solid lines) and B (dashed lines) for different differential velocities of an infalling galaxy. 
\label{RP_cl_model}}
\end{figure}

\begin{table}
\begin{tabular}{lccccc}
\hline
Model 		&$\sigma_{cl}$  	& $R_{200}$  	& $\beta$ 	& $\rho_0$ 	& $R_c$  \\
name		&(km~s$^{-1}$) 		&  (Mpc)   	& 		& ($10^{-23}$Kg~m$^{-3}$) & (kpc) \\
\hline
Cluster A 	& 593		& 1.55		& 0.5	  	& 6.69		& 13	\\
Cluster B   	& 982		& 2.37		&0.5	  	& 4.30		& 82	\\
\hline 
\end{tabular} 
\caption{Properties of the model clusters: low-mass cluster (A) and high-mass cluster (B).
The columns are: cluster, velocity dispersion, $R_{200}$, $\beta$ parameter from equation~\ref{eq:beta}, central gas density $\rho_{0}$. References for Cluster A: \citet{Mei2007,McLaughlin1999, Ferrarese2012}. Reference for Cluster B: \citet{Chen2007}.}
\label{cluster_models}
\end{table}

\section{Modelling ram-pressure stripping analytically}
\label{sec:rps_model}

In the following we present models of ram-pressure stripping intensity based on two different cluster models and two different galaxy models. 

Ram-pressure by the ICM was defined by \citet{GunnGott1972} as: 
\begin{equation}
P_{ram}=\rho_{_{ICM}} \Delta v_{cl(3D)}^2
\label{eq:pram}
\end{equation}

where $\Delta v_{cl(3D)}$ is the differential 3D velocity of the galaxy with respect to the cluster centre, and $\rho_{_{ICM}}$ the density of the ICM, which decreases with clustercentric radius.

When $P_{ram}$ exceeds the anchoring self-gravity provided by the galaxy ($\Pi_{gal}$), the galaxies' gas is stripped.

In the following, we create general cluster ($\S$\ref{sec:Pram}) and galaxy ($\S$\ref{sec:PI}) models that can conveniently be used in other studies to predict the amount of gas stripped ($\S$\ref{sec:rstripped})  and define ram-pressure stripping zones in position vs. velocity phase-space diagrams ($\S$\ref{sec:rps_pps}).

 \begin{figure}
\centering
\includegraphics[width=0.48\textwidth]{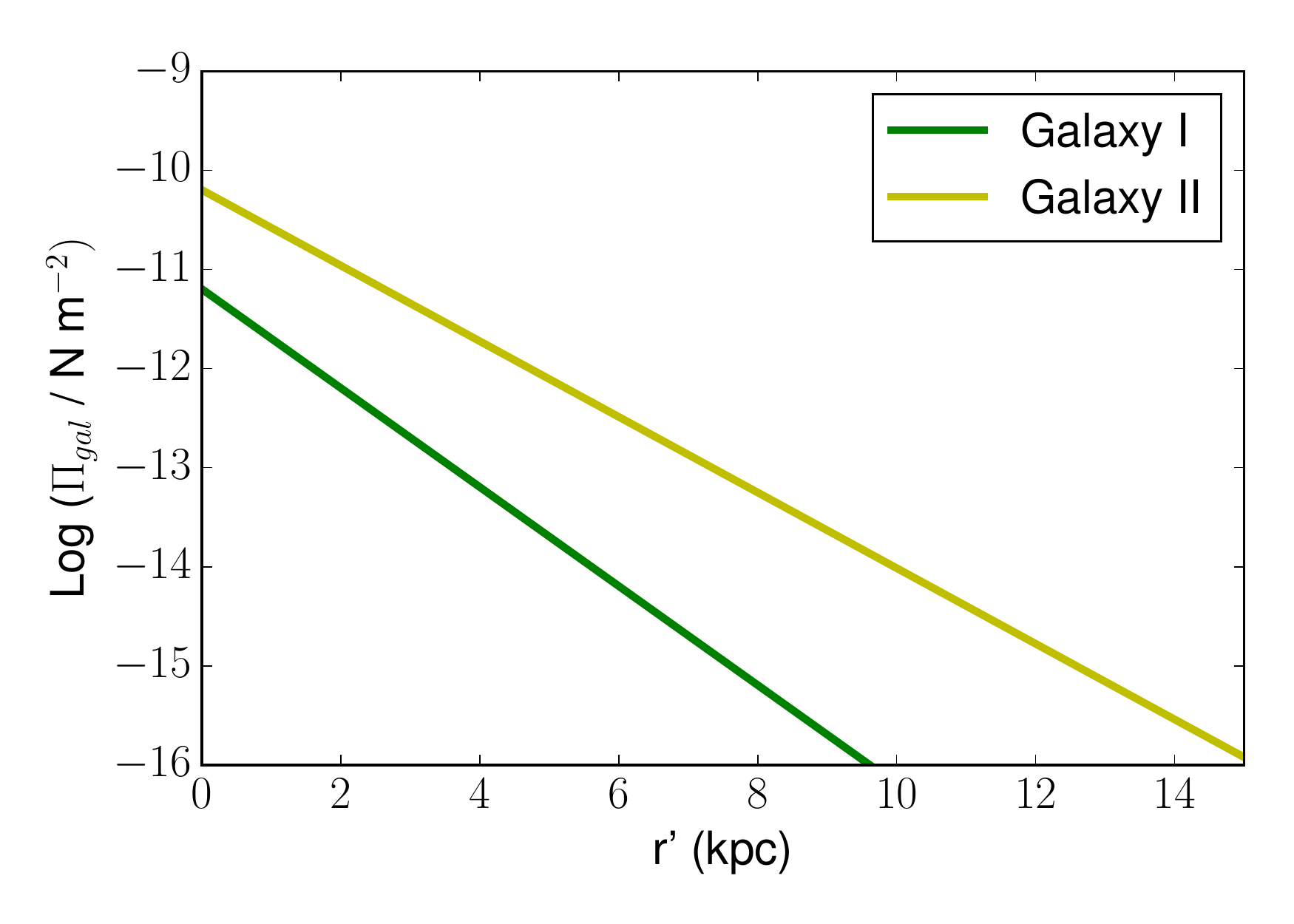}
\caption{The anchoring force as a function of radial distance from the galaxy centre for two model galaxies (I and II, as indicated). 
\label{gal_PI_r}}
\end{figure}

\begin{table}
\begin{tabular}{lcccc}
\hline
Model & $ \log M_{\bigstar}$  	& $M_{gas}/M_{\bigstar}$ & $R_d$  & $\Pi_{gal(r=0)}$  \\
name & ($M_{\odot}$) 		&  			 & (kpc)  & ($N m^{-2}$) \\
\hline
Galaxy I  	& 9.2		& 0.7	& 1.4	& $ 9.2 \times 10^{-12}$ \\
Galaxy II  	& 10.1		& 0.3	& 2.2	& $ 8.9 \times 10^{-11}$ \\
\hline 
\end{tabular} 
\caption{Properties of the model galaxies: low-mass galaxy (I), and high-mass galaxy (II). Columns are: stellar masses ($M_{\bigstar}$), gas fractions \citep[include $HI$ and $H_2$, from][]{Popping2014}, disk scale-length ($r_d$,  mean values of the OmegaWINGS galaxies in the low- and high-mass bins shown in the right-hand panel of Figure~\ref{cl_properties}), and the resulting anchoring force of the galaxy ($\Pi_{gal}$, computed from equation~\ref{eq:pigal}).}
\label{galaxy_models}
\end{table}

\subsection{Cluster models}
\label{sec:Pram}

The ICM gas density profile of clusters is typically parametrized with a  $\beta$-model \citep{Cavaliere1976}, as:
\begin{equation}
\rho_{_{ICM}}(r_{cl({3D})})=\rho_0 \left[1+\left(\frac{r_{cl({3D})}}{R_c}\right)^2\right]^{-3\beta/2}
\label{eq:beta}
\end{equation}

where $\rho_{0}$ is the gas density at the cluster center, $R_{\mathrm{c}}$ the core radius, and $r_{cl({3D})}$ the 3D distance from the cluster centre. 

We use two cluster models that seek to cover the extremes of the cluster population. Their properties are listed in Table~\ref{cluster_models}. 
Cluster A represents low-mass cluster population. Hence, we use the ICM of the Virgo cluster as reference as it is a well studied  in the context of ram-pressure stripping \citep[e.g][]{Vollmer2001,Chung2009,Boselli2011,Ferrarese2012,Yoon2017}, and widely referenced in galaxy evolution studies. 
Cluster B instead represents high mass clusters. In this case we used Abell 85 (A85) as reference as it is one of the most massive clusters in WINGS and it is  also well studied \citep[e.g.][]{Kempner2002,Durret2005,BravoAlfaro2009,Aguerri2017}. 

Figure~\ref{RP_cl_model} shows $P_{ram}$ as a function of distance from the core of Virgo (Cluster A, dashed line) and A85 (Cluster B, solid line), computed using equations \ref{eq:pram} and \ref{eq:beta}) for different galaxy velocities. 

Note that, although at the cluster centres $P_{ram}$ is similar in both clusters, at increasing $r_{cl({3D})}$ they quickly start to differ by almost an order of magnitude,  and if we consider the typical velocities of galaxies in each cluster ($\sigma_{cl}$ from $\sim 400$ to $\sim1300$ km~s$^{-1}$), the difference can become even larger.

\subsection{Galaxy models}
\label{sec:PI}

To assess whether the cluster can strip the gas in an infalling galaxy it is necessary to have an estimate of the orbit and also the galaxy's anchoring (restoring) force or self-gravity.

The galaxies's anchoring force, $\Pi_{gal}$, can be modelled assuming the form: 

\begin{equation}
\Pi_{gal}(r')=2 \pi G \Sigma_{\rm s}\Sigma_{\rm g} 
\label{eq:pigal}
\end{equation}

where $\Sigma_{\rm s}$ and $\Sigma_{\rm g}$ are the density profiles of the stellar and gaseous disks respectively. 

Given that most jellyfish galaxies are gas-rich late-type galaxies \citep[although see the peculiar cases of stripping presented in][and Moretti et al. submitted]{Sheen2017}, we assume an exponential (disk) profile for the density of both the gas and the stars:

\begin{equation}
\Sigma(r')=\Sigma_0 e^{-r'/R_d}
\label{sigma0}
\end{equation}

where $r'$ is the radial distance from the centre of the galaxy,   $R_d$ is the disk scale length, $\Sigma_0=M_d / (2 \pi R_d^2)$,  
and $M_d$ is the mass in the disk.

In order to create generic disk galaxy models we explored the properties of the disky spiral galaxies in WINGS/OmegaWINGS, 
plotted in the right hand side of Figure~\ref{cl_properties}. We distinguished field (blue) from cluster (red) galaxies and found no strong segregation in either mass or size. 
Since the galaxies are 
disk-dominated we assume that the stellar disk ($M_{d,\bigstar}$) is equivalent to the total stellar mass of the galaxy. We discuss the impact of the inclusion of a bulge component in Section~\ref{sec:caveats}.

Another parameter we had to assume was the gas fraction. We used as reference the work of \citet{Popping2014}, who  show the variation of HI and $H_2$ content of disk-dominated galaxies (B/T$<$0.4) as a function of stellar mass. In the mass range $10^{8}<M_{\bigstar}<10^{11.5}M_{\odot}$  the fraction of HI mass relative to stellar mass varies between 150 to 5\%. For $10^{10}M_{\odot}$ galaxies the HI gas fraction is $\sim 20\%$. To this fraction we need to add the $H_2$ gas component, which is less abundant and varies much less with $M_{\bigstar}$ (from 10 to 4\% in the abovementioned mass range).

 \begin{figure}
\centering
\includegraphics[width=0.48\textwidth]{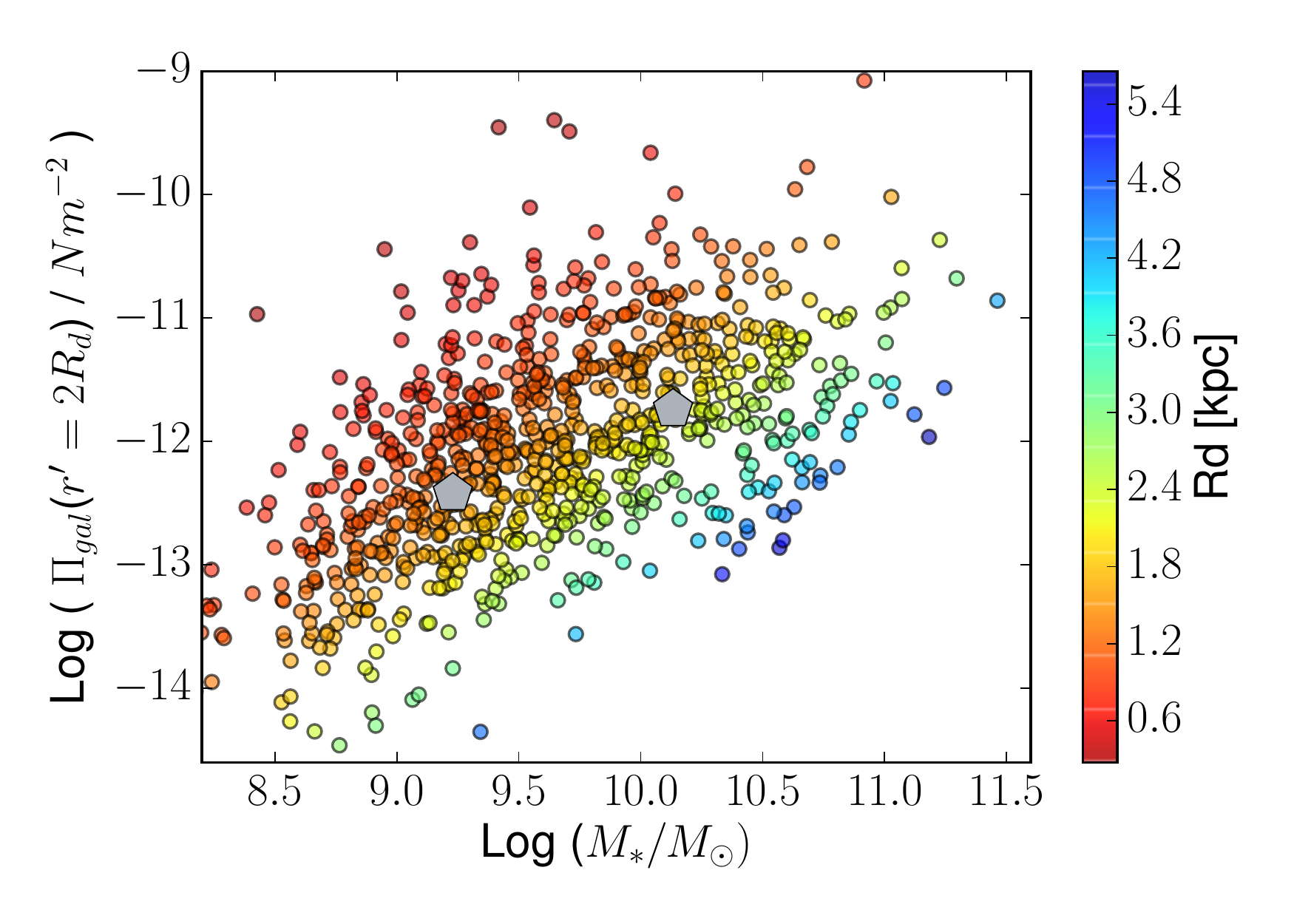}
\caption{The anchoring force at $r^{\prime} = 2\times R_{d}$ as a function of stellar mass for all spiral galaxies in our (cluster plus field) sample. 
The grey  pentagons correspond to the the model galaxies I and II. 
\label{gal_PI}}
\end{figure}

We assume that the extent of the gas ($r_{d,g}$) is larger than the size of the stellar disk ($r_{d,\bigstar}$) by a factor of $=1.7$. This value corresponds to the HI to optical radius of non HI-deficient spiral galaxies in Virgo \citep{Cayatte1994} 

Having a good characterization of the galaxies we can now compute $\Pi_{gal}$. We do this for two idealized galaxies (Galaxy I and II) that represent the mean of the low ($< 5 \times 10^{9}M_{\odot}$) and high ($> 5 \times 10^{9}M_{\odot}$) mass galaxies in our galaxy sample (see right hand panel of Figure~\ref{cl_properties}). The properties of the model galaxies are listed in Table~\ref{galaxy_models}, and the resulting $\Pi_{gal}$ profile as a function of $r$ is shown in Figure~\ref{gal_PI_r}.

For completeness and future reference, Figure~\ref{gal_PI} further shows $\Pi_{gal}$ computed at $r=2\times R_d$ as a function of stellar mass for WINGS/OmegaWINGS galaxies, colour-coded by disk scale length. 
The figure highlights the interconnected role of stellar mass and size in the estimation of the restoring force of a galaxy. It is interesting to notice that a dwarf galaxy with $M_{\bigstar} = 1\times 10^9 M_{\odot}$ can in principle have the same central restoring force as a galaxy 100 times more massive ($M_{\bigstar} = 1\times 10^{11} M_{\odot}$) if the sizes are very different.  Galaxies with higher central densities (red symbols) have stronger anchoring pressure.

\subsection{Computing the amount of stripped gas}
\label{sec:rstripped}

Having a model of the galaxy ($\S$\ref{sec:PI}) and the cluster ($\S$\ref{sec:Pram}) that it is infalling into, it is possible to estimate the amount of stripping (total gas mass lost by the galaxy) 
by directly comparing $P_{\rm ram}$ (Figure~\ref{RP_cl_model}) with $\Pi_{\rm gal}$ (Figure~\ref{gal_PI_r}). We summarize the steps below: 
\begin{enumerate}
 \item Obtain the position and velocity of the galaxy (or galaxies) within its host cluster. These are the phase-space coordinates.  Note that because these quantities come from photometric and spectroscopic observations of the cluster, one can only measure projected distances ($r_{cl}$) and line-of-sight velocities ($\Delta v_{cl}$), which are lower limits to the real (3D) values (see discussion in Section~\ref{sec:caveats}). 
 \item Using the (i), compute $P_{\rm ram}$ from equations~\ref{eq:pram} and~\ref{eq:beta} if $\rho_{\rm ICM}$ is known (from X-ray observations of the cluster of interest). Alternatively, a scaled estimate can be inferred from the relations displayed in Figure~\ref{RP_cl_model}. 
 \item Compute the anchoring force of the galaxy, $\Pi_{\rm gal}$ from equations~\ref{eq:pigal} if the galaxy mass, size and morphological type is known, or get a scaled estimate from the relations plotted in Figure~\ref{gal_PI_r} and Figure~\ref{gal_PI}.  
 \item Compare the $P_{\rm ram}$ obtained in (ii) with the radial profile of $\Pi_{gal}$ for the galaxy obtained in (iii), and identify the radius  $r_t$ within the galaxy disk at which $P_{\rm ram} = \Pi_{gal}$. Because ram-pressure strips the disk gas outside-in, $r_t$ corresponds to the radius of stripping, or  ``truncation'' radius. 
 \item From $r_t$ we can further compute the remaining gas mass in the galaxy following: 

\begin{equation}
f=1+\left[e^{-r_t/R_d}\left(\frac{-r_t}{R_d} -1\right)\right] \label{eq_tr}
\end{equation}
\end{enumerate}

This equation is directly derived from the mass distribution of an exponential disk assuming the remaining gas is enclosed inside the truncation radius ($r'< r_t$).

\begin{figure}
\centering
\includegraphics[width=0.48\textwidth]{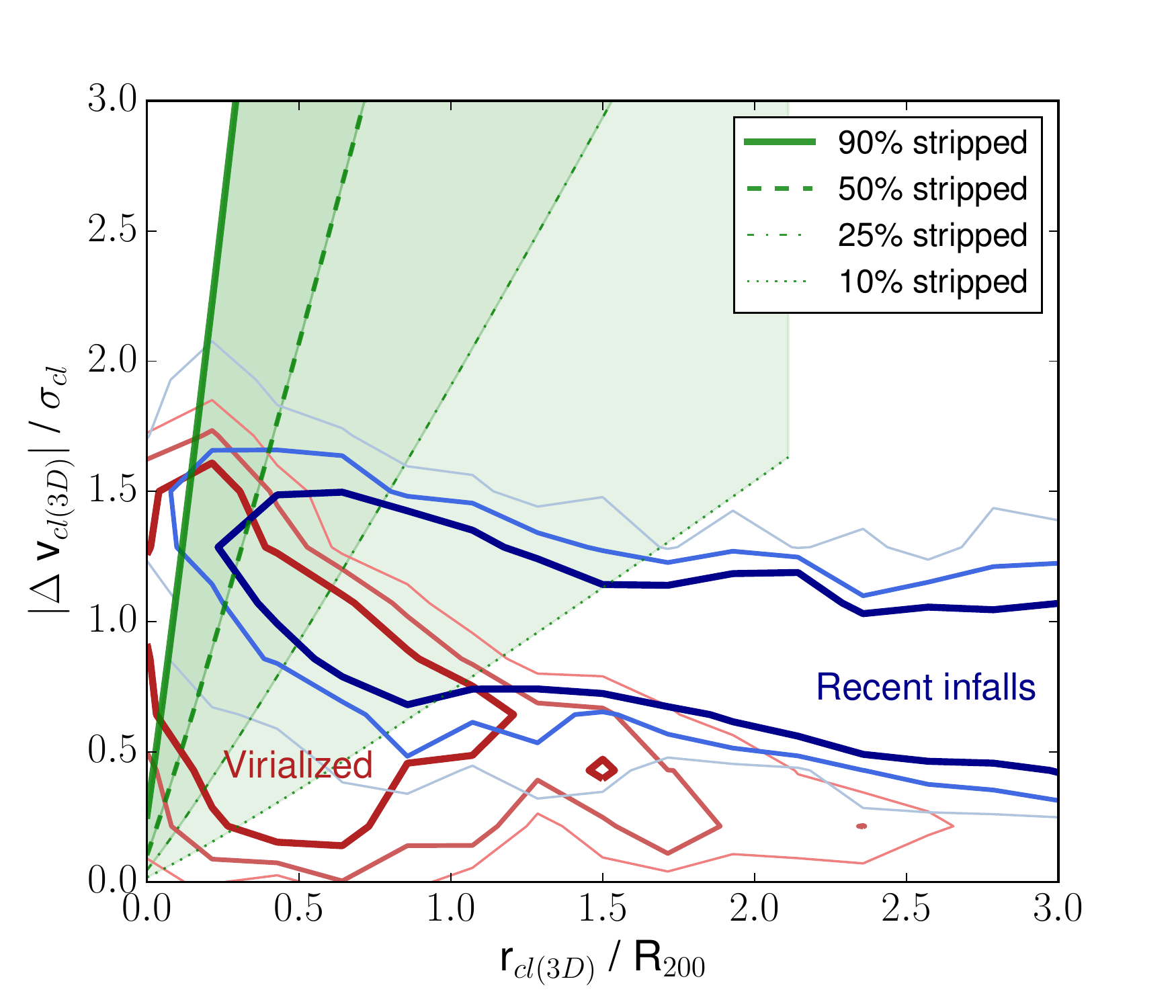}
\caption{The position vs. velocity phase-space diagram of 15 simulated group and cluster galaxies (mass range from $0.5 \times 10^{13}$ to $1 \times 10^{15}M_{\odot}$) from \citet[][private communication]{Rhee2017}, considering multiple lines of sight, separated into ``virialized'' (entered the cluster $>4$ Gyr ago; red contours) and ``Recent infalls'' (falling towards the cluster for the first time or recently entered the cluster $<2$ Gyr; blue contours). A galaxy is considered to enter the cluster when it has crossed $R_{200}$ for the first time.  The contours enclose 1000, 2500, and 5000 particles from lighter to darker colours respectively. The axes have been normalized by cluster size ($R_{200}$) and cluster velocity dispersion ($\sigma_{cl}$) to allow the stacking of 15 simulated clusters with masses between $10^{13}$ and $10^{15} M_{\odot}$. 
The green area indicates the region where ram-pressure by the ICM is able to strip 10\% (dotted line) to 90\% (solid lines) of the total gas mass of an low-mass galaxy (model galaxy I) falling in a massive cluster (model cluster B). 
For details see Section~\ref{sec:rps_model}, and Tables~\ref{cluster_models} and~\ref{galaxy_models}.  \label{PPS_clusters} }
\end{figure}

\subsection{Ram-pressure stripping in phase-space}
\label{sec:rps_pps}

Position vs. velocity phase-space diagrams are very useful tools to study the orbital histories of galaxies in clusters. Moreover, these diagrams can be utilized to study cluster processes such as ram-pressure stripping \citep{HernandezFernandez2014,Jaffe2015, Yoon2017} or tidal mass loss \citep[][]{Rhee2017}. 

Figure~\ref{PPS_clusters} shows schematically the position vs. velocity phase-space location of simulated cluster galaxies from \citet{Rhee2017},  separated by time since infall and normalized by cluster size and velocity dispersion. Infalling galaxies will approach the cluster centre gaining velocity, so they are  located within the blue contours of the figure. After the galaxy has passed pericentre, it will inevitably ``bounce'' back in phase-space but with lower velocity due to the combined effect of violent relaxation, dynamical friction, and the growth of the cluster. 
For this reason, galaxies that have been in the cluster for over a cluster crossing time will tend to gather near the centre of the cluster at lower velocities (red ``virialized'' region). Note that the ``virialized'' and ``Recent infalls'' regions, although fundamentally different, have some overlap at high velocities in the inner part of the cluster ($r_{cl(3D)} < R_{200}$).

From the modelling described in Sections~\ref{sec:Pram},~\ref{sec:PI}, and~\ref{sec:rstripped} it is possible to construct regions in phase-space where ram-pressure stripping is at play. 
This is shown in the green ``cone'' of  Figure~\ref{PPS_clusters}, that represents as an example, the region where a low-mass galaxy (model Galaxy I) would be incrementally stripped (losing from 10 to 90\% of its gas) by ram-pressure as it falls into a massive cluster (Cluster B). The different green curves further show the increasing effect of ram-pressure with decreasing clustercentric distance and increasing velocities, in accordance with equation~\ref{eq:pram}. As more gas gets stripped (indicated by the percentages of total gas mass lost in the legend), the truncation radius will inevitably become smaller. Note that the green lines do not reach zero velocity because ram pressure intensity depends on velocity squared.

It is important to highlight that the stripping region in phase-space depends on cluster mass and galaxy mass/size, with lower-mass, less concentrated galaxies being more vulnerable to stripping than massive galaxies. Ram-pressure stripping ``cones'' generated by different clusters for different galaxies are shown in Figure~\ref{PPS_jellys_byclM} and will be discussed in Section~\ref{sec:orbits}.

\subsection{Caveats}
\label{sec:caveats}

There are several caveats to the simplistic ram-pressure stripping model presented in this section. We summarize the most important ones here:  

\begin{itemize}
 \item  Scatter introduced by projection effects.  Observationally, one measures projected radius in the plane of the sky and line-of-sight velocities. The 3D radius scales with the projected radius as $r_{cl}=cos(\varphi)r_{cl(3D)}$, where $\varphi$ is the projection angle, that can vary between 0 and $\pi/2$. On average, $r_{cl} = \frac{2}{\pi} r_{cl(3D)}$ but there is considerable scatter. The projected velocity is also a lower limit to the 3D velocity by a factor of $\sqrt{3}$ on average.  

 Despite the known projection effects, several works using cosmological simulations have shown that it is possible to separate the oldest from the most recently accreted cluster members in projected phase-space. Galaxies with intermediate time-since-infall (including backsplash galaxies) are harder to distinguish, as they overlap in phase-space with both the virialized and the recent infall galaxies \citep[][]{Oman2013,Jaffe2015,Rhee2017}. However,\citet{Yoon2017} recently showed that it is possible to identify the backsplash population combining the phase-space location of cluster galaxies with detailed HI observations, which they used as a proxy for the galaxy's stripping stage and thus time since infall.

 \item  The analysis presented in this section applies to disk galaxies. For non-disk galaxies the calculation of the restoring force should be altered. 
  
 \item  The analysis also assumes the clusters are regular (dynamically relaxed), with a spherical $\beta$-profile for the ICM density (Equation~\ref{eq:beta}). For clusters with significant substructure it should be noted that the intensity of ram-pressure can be enhanced at places and inhomogeneous across the cluster and that the orbits (and thus velocities) of infalling galaxies can also be altered  \citep{Owers2012,Vijayaraghavan2013,Jaffe2016,McPartland2016}. We note however that most of the clusters considered in this paper have a regular distribution of galaxies (both spatially and in velocity). Only a fraction ($\sim 1/3$) of clusters show signs of galaxy substructure, but the contribution of such substructures are expected to become less significant when we stack all the clusters together in phase-space (c.f. Section~\ref{sec:jelly_pps}). Nevertheless, owing to the known importance of cluster and group mergers in the evolution of galaxies, we will use our sample of clusters to study the effect of cluster substructure and cluster mergers in the formation of jellyfish galaxies in a dedicated follow-up study.
 
 \item  The \citet{GunnGott1972} approximation assumes a face-on interaction between the galaxy and the ICM. Although simulations show that the inclination of the galaxy with respect to the cluster (i.e. the wind angle) can lead to complex behaviour \citep[e.g.][]{Abadi1999,Vollmer2001}, it has been shown that the amount of stripping is only significantly altered in the case of edge-on interactions \citep{Roediger2006}, which are very rare.

 \item We assume the disk galaxies are not stripped in any way before entering the cluster, which might not be the case if there were any pre-processing in lower density environments \citep[see e.g.][]{Jaffe2016}. 

 \item Finally, we note that because our galaxy models assume a pure disk profile, they may have underestimated the true anchoring force by neglecting the contribution of the dark matter halo and bulge. In principle this could result in an overestimation of the gas stripping. We estimate the additional contribution of the bulge and dark matter halo using analytical models. Our test models consist of an \citet[][NFW]{NFW96} halo, a \citet{Hernquist1990} profile for the bulge, and the same exponential disks of gas and stars of Galaxy I and II. 
 We measure the restoring force for gas within $\sim 10 \times R_d$, near the plane of the galaxy, and we consider a wide range of halo concentrations ($c=10$ to $30$). We find that the contribution of the bulge and halo can be significant, but only in the very inner disk (at $< 0.5 R_d$).
At $2 R_d$, the addition of both the bulge and halo increases the anchoring force by no more than a few percent, which we calculate has only a very mild impact on predicted truncation radii.

\end{itemize}

\begin{figure*}
\centering
\includegraphics[width=1.0\textwidth]{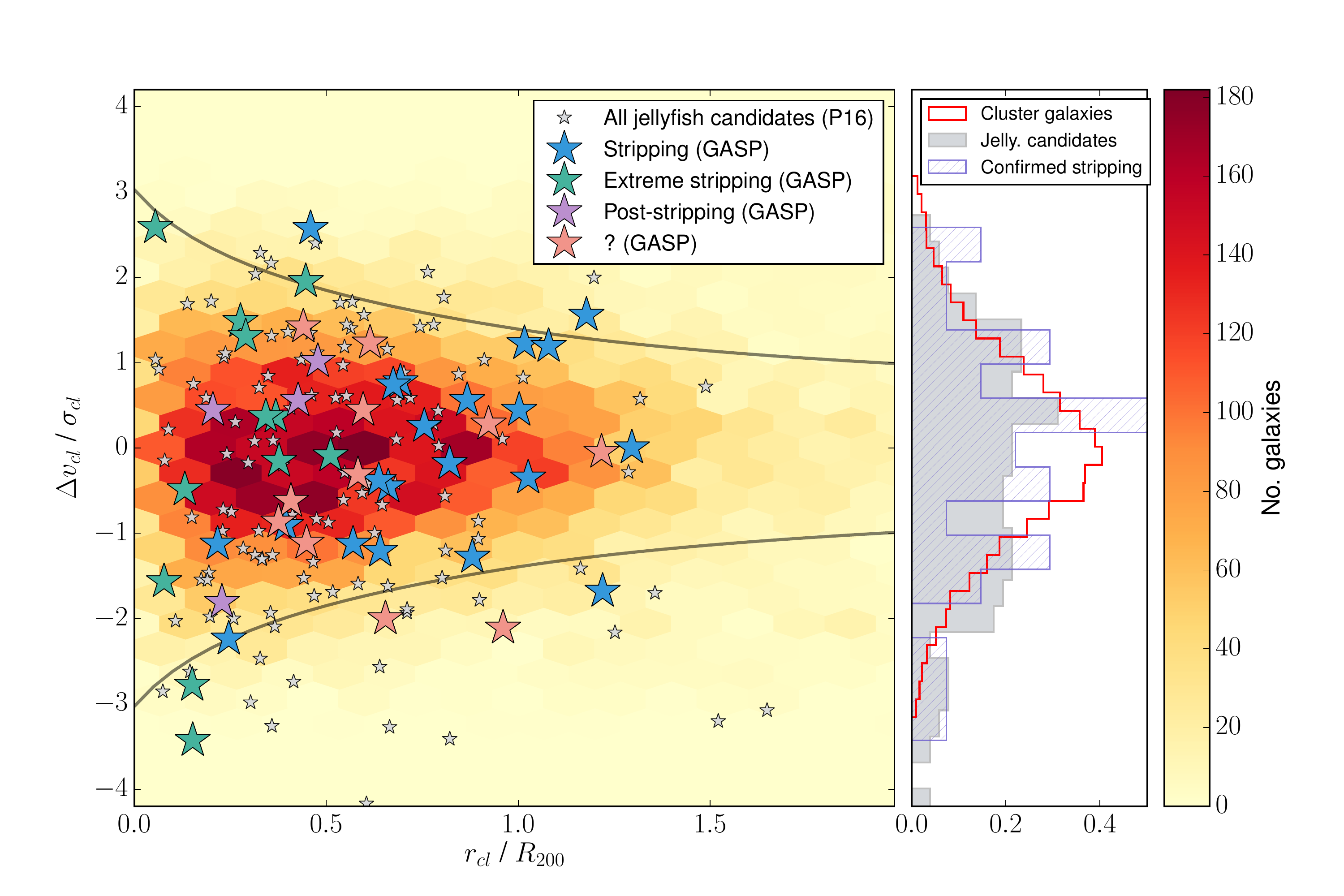}
\caption{The location in projected position vs. velocity phase-space of all the jellyfish galaxy candidates from P16 (small gray stars) and the ones observed with MUSE by GASP so far (larger colored stars), separated by stripping stage as indicated. The background shows the distribution of all WINGS/OmegaWINGS clusters with spectroscopic completeness $> 50$\% stacked together (orange colorbar). The gray curve corresponds to the 3D (un-projected) escape velocity in a NFW halo with concentration $c=6$ for reference. 
Note that, contrary to the absolute velocities plotted in the phase-space diagram of Figure~\ref{PPS_clusters}, the velocity axis in this plot has positive and negative values. 
To the right of the phase-space diagram, a plot shows the velocity distribution of the overall cluster population of galaxies (open red histogram), all the jellyfish candidates from P16 (filled grey histogram), and the galaxies observed by GASP that are confirmed stripping cases (i.e. ``Stripping'', ``Extreme stripping'', and ``Post-stripping; dashed blue histogram) at $r_{cl} < R_{100}$. All histograms have been normalized to unity for comparison. 
\label{PPS_jellys}}
\end{figure*}

\begin{figure*}
\centering
\includegraphics[width=0.45\textwidth]{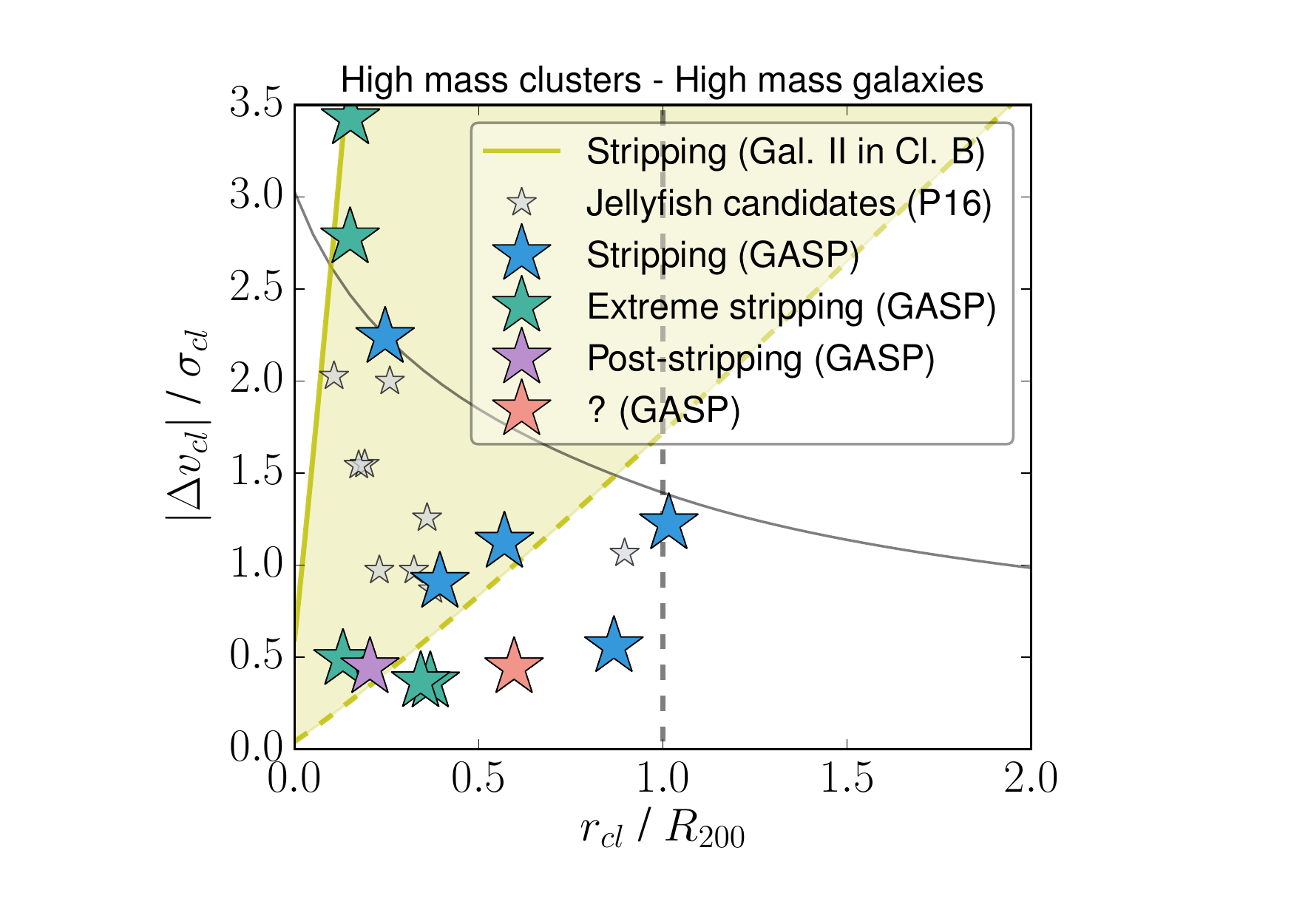}
\includegraphics[width=0.45\textwidth]{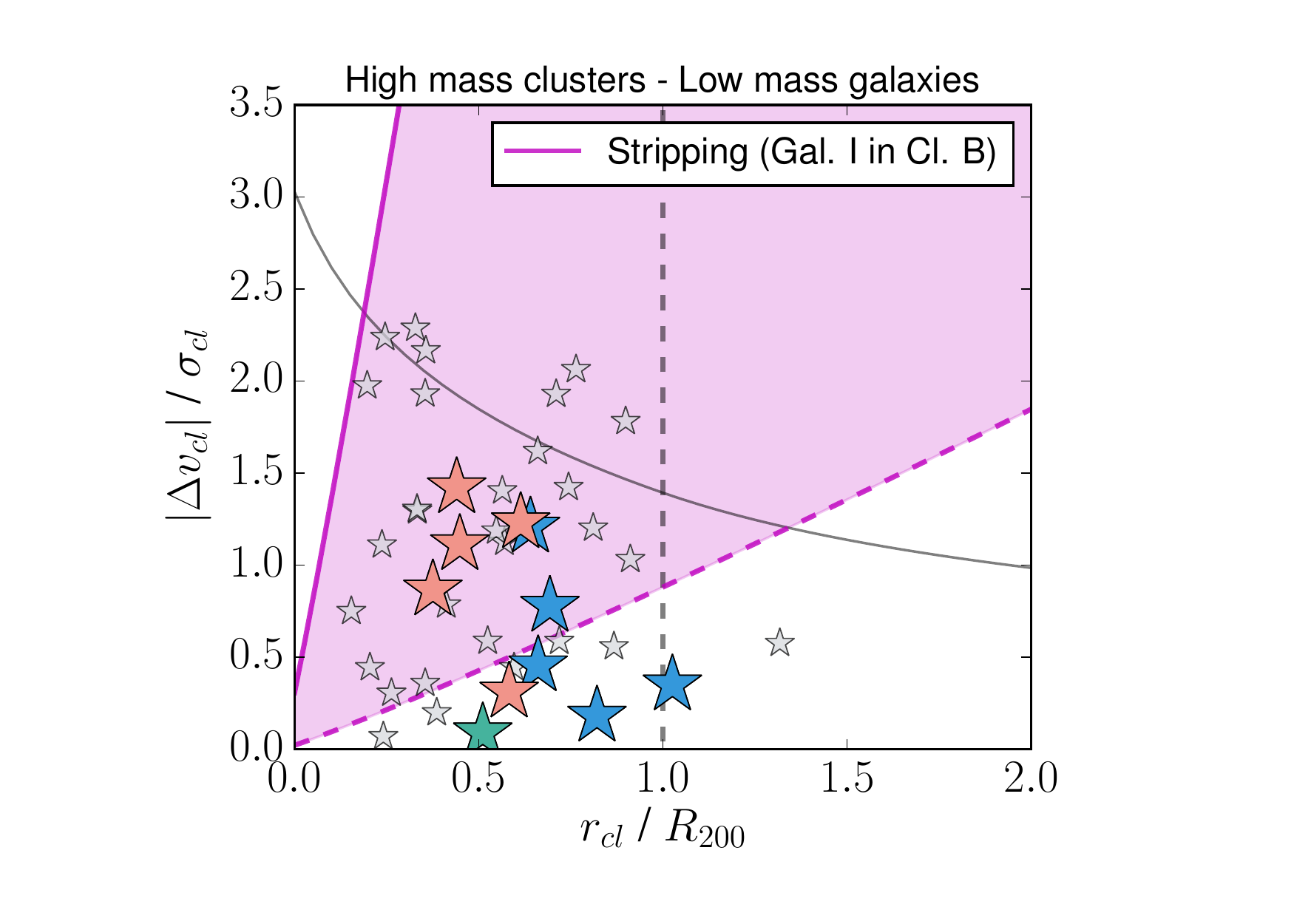}
\includegraphics[width=0.45\textwidth]{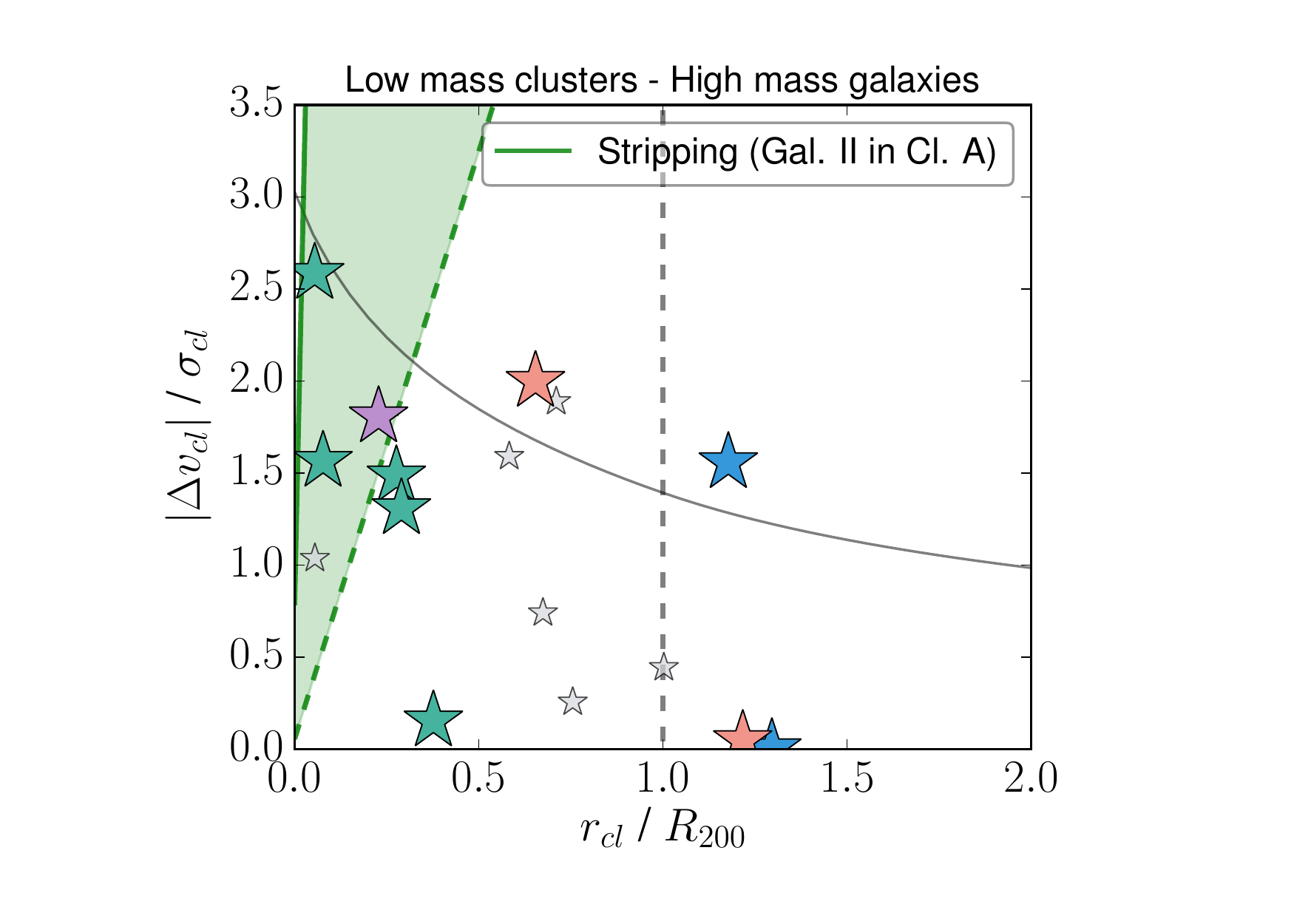}
\includegraphics[width=0.45\textwidth]{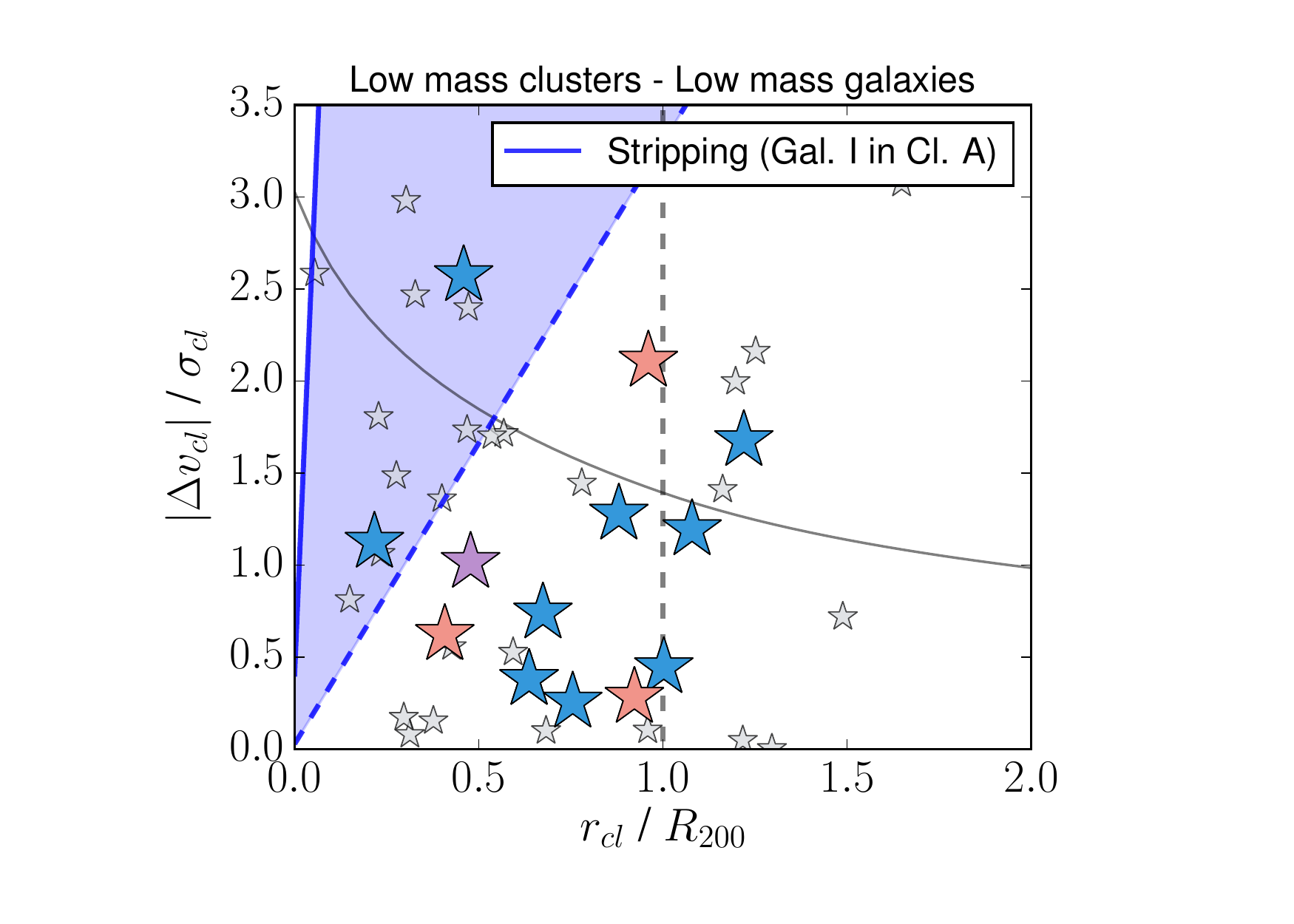}
\caption{The location in position vs. velocity phase-space of jellyfish galaxies in high mass clusters (top panels; $\sigma_{cl}>750 km/s$) and low mass clusters (bottom panels; $\sigma_{cl}<750lkm/s$)) separated in two bins of stellar mass: $>5 \times 10^{9}M_{\odot}$ on the left panels and  $<5 \times 10^{9}M_{\odot}$ on the right panels. As in Figure~\ref{PPS_jellys}, grey stars correspond to the jellyfish sample of P16 while the bigger stars are the confirmed jellyfish galaxies from GASP. 
Lines of different intensities of ram-pressure  are indicated in each case: the dashed coloured lines indicate 10\% stripping of the total gas mass, while the solid line corresponds to 90\% stripping. The grey curve corresponds to the escape velocity in a NFW halo. The vertical dashed line indicates $r=R_{200}$, which is roughly the extent to which all clusters used have a high spectroscopic completeness. \label{PPS_jellys_byclM}}
\end{figure*}

\section{The orbital history of jellyfish galaxies in clusters}
\label{sec:orbits}

In the following, we use projected position and line-of-sight velocity information of the cluster jellyfish galaxies to infer the time since infall into the cluster, the eccentricity of the orbits, and the intensity of ram-pressure stripping exerted by the ICM, modeled in  Section~\ref{sec:rps_model}.

\subsection{Jellyfish galaxies in phase-space}
\label{sec:jelly_pps}

Figure~\ref{PPS_jellys} shows the projected phase-space distribution of all the cluster members in our galaxy sample (orange background) stacked together in a ``master cluster''. 
The densest part of the plot (darker orange) corresponds to the ``virialized'' part of the clusters. 
The location of jellyfish galaxies in phase-space is highlighted with stars on top of the overall cluster population. Smaller symbols correspond to jellyfish candidates from P16 while larger symbols represent confirmed jellyfish from GASP. 
Given the large size of the WINGS/OmegaWINGS cluster galaxy sample, we limited the cluster sample to the clusters with completeness $>$50\% to reduce biases, but plotted jellyfish galaxies in all clusters to increase the number count. We checked however that the distribution of jellyfish galaxies restricted to clusters with spectroscopic completeness $>$50\% in phase-space is very similar to the distribution of all jellyfish galaxies. 
In the region where there is higher spectroscopic coverage for all clusters ($r<R_{200}$) the incidence of jellyfish relative to the overall population of cluster galaxies peaks 
near the reference escape velocity curve,  avoiding the virialized part of the cluster where most of the cluster members that are not currently undergoing stripping are. The difference in the distributions can be quantified with a 2D Kolmogorov-Smirnov (KS) test, that gives a very low probability (0.6\%) that the samples of stripping (P16 jellyfish; stars) and non-stripping  (WINGS/OmegaWINGS; colourbar) cluster galaxies within $r_{cl}<R_{200}$ are extracted from the same parent distribution.

Focusing on the GASP galaxies, it is noticeable how the jellyfishes with longest tentacles of stripped material (``Extreme stripping''; green stars) are all located within $0.5 \times R_{200}$ and many are at $|\Delta v|/\sigma >1$, which results in a broad velocity distribution. Their location coincides with the most favorable conditions for ram pressure stripping within the cluster, which requires a small projected cluster-centric radius, and high absolute differential velocity with respect
to the cluster redshift (see $\S$\ref{sec:rps_model}). 
In fact, in the first papers from the GASP series \citep{Poggianti2017a,Bellhouse2017,Fritz2017,Gullieuszik2017} we have carefully compared the truncation radius measured from the $H_{\alpha}$ maps with the predictions from the position of three of these galaxies in phase-space. We found in all cases a good agreement between the predicted and measured stripping. In particular, we find that the galaxies have lost a significant fraction ($\gtrsim15$, 50, and 40\%) of their total gas mass during their first infall into the cluster.  Moreover,  hydrodynamical simulations of one of the galaxies \citep[JO204; ][]{Gullieuszik2017} further confirm the estimated amount of stripping.

The galaxies with milder signatures of stripping (``Stripping''; blue stars in Figure~\ref{PPS_jellys}), unlike the``extreme stripping'' cases, are mostly located beyond $>0.5 \times R_{200}$, but they share the elevated velocities and wide velocity distribution with respect to the general cluster population (see histogram in the right-hand panel of Figure~\ref{PPS_jellys}). 
The distribution of both populations of stripped galaxies in phase-space coincides remarkably well with the region of ``recent infallers'' defined from cosmological simulations in \citet[][]{Rhee2017} and shown in Figure~\ref{PPS_clusters}.  However, the distribution of galaxies undergoing moderate ``Stripping'' is notably different than that ``Extreme stripping'' galaxies. In fact, the K-S probability that they are drawn from the same parent distribution is 0.026\%. 

Interestingly, the galaxies with heavily truncated gas disks (``Post-stripping''; purple stars) are near the centre of the cluster, 
which indicates they had an orbit which brought them close to the core of the cluster, where ram-pressure is most intense. 
Although so far GASP has only observed 4 ``Post-stripping'' galaxies, their moderate velocities could be interpreted as an indication of deceleration post-pericentric passage. 

Finally, the galaxies with inconclusive gas morphologies (``?'', orange stars) share common location with the moderately stripped galaxies, not ruling out the effect of ram-pressure. 

To assess the intensity of ram-pressure that the galaxies could be experiencing, we split the sample of GASP galaxies in 4 bins, according to their stellar masses, and the masses of the clusters. We consider low mass galaxies all the galaxies with $M_{\bigstar} < 5 \times 10^{9} M_{\odot}$ and high-mass galaxies $M_{\bigstar} > 5 \times 10^{9} M_{\odot}$.  Likewise, we separate cluster by mass, splitting them in velocity dispersion at $750$ km s$^{-1}$. Figure~\ref{PPS_jellys_byclM} shows the phase-space distribution of jellyfish in those 4 bins. In each case, the area where ram-pressure stripping is at play (from 10\% to 90\% stripping) is indicated with a shaded colored area. 
The upper-right panel corresponds to the case of most efficient stripping, while the lower-left panel corresponds to the least efficient stripping, with gas loss only occurring in a limited range of distances and velocities. This is partly reflected in the larger number of jellyfish found in the top-right panel with respect to the lower-left, although there might be selection effects related to the impact of the viewing angle and stripping stage in the identification of jellyfish galaxies.

GASP galaxies undergoing ``Extreme stripping'' (green stars in Figure~\ref{PPS_jellys_byclM}) are high mass galaxies in both low- and high-mass clusters. With only one exception, these galaxies are inside or  near the zone where ram-pressure is estimated to be most intense.  It is important to keep in mind that the plotted velocities are lower-limits to the real 3D velocities of the galaxies, and that these particular galaxies display long tails of stripped gas, which implies that a large component of their velocities are in the plane of the sky. In other words, it is likely that these galaxies would move up in the velocity direction in real 3D phase-space. Taking this into consideration we conclude that galaxies classified as ``Extreme stripping'' from their $H_{\alpha}$ morphologies are consistent with being subject to intense ram-pressure close to pericentric passage (i.e. peak stripping). This result was partially shown in Fig. 4 of \citep{Poggianti2017b},  where among the 7 GASP galaxies with the strongest signatures of stripping observed with MUSE by then, the majority were found to host an AGN. All of these galaxies are located in the area of phase-space where ram pressure stripping is the strongest, supporting the hypothesis that ram pressure can induce AGN activity. Ram-pressure stripping simulations by \citet{TonnesenBryan2012} show that the star-formation rate in the stripped tails depends on the amount of gas stripped and the ICM pressure. It is thus possible that only  massive galaxies were able to keep their gas for long enough to still be gas-rich by the time they approach pericentre. If they pass close to the cluster core (as our ``Extreme stripping'' galaxies do),  the gas that is stripped will be able to form more stars. Another possibility is that massive galaxies will show bright (detectable) optical tails for longer than low-mass ones. We will investigate these possibilities with the full GASP sample in future studies.

It is interesting to note that many of the jellyfish galaxies with less extreme signs of stripping  (``Stripping'', blue stars in Figure~\ref{PPS_jellys_byclM}) are low-mass galaxies in low-mass  clusters located outside the high-intensity stripping zone. As explained in Section~\ref{sec:caveats}, the main caveat in the interpretation of the distribution of galaxies in phase-space is the effect of projection, which makes the measured distances and velocities lower limits to the real values. 
The one-sided tails of jellyfish galaxies indicate that they are moving at least partially along the plane of the sky, which in turn means that their projected velocities are likely lower than their true value. 
It is not trivial to infer the 3D velocity of observed cluster galaxies but it is reasonable to argue that, on average, the measured (line-of-sight) velocities of jellyfish galaxies are more likely to be an underestimation of the real (3D) value in comparison to other cluster galaxies.
Considering this, it is possible that, after removing projection offsets, many of these ``Stripping'' jellyfish outside the high-intensity stripping zone in projected phase-space would be inside this zone if de-projected. 
Alternatively, it is possible that  some of the jellyfish galaxies could reside in clusters with significant sub-structure (e.g. the Shapley supercluster), where  multiple cluster centres are present and/or ram-pressure is likely enhanced outside the cluster core. 
Although in our sample most clusters are not undergoing mergers, if we exclude the most unrelaxed clusters from the analysis, we find that the remaining jellyfish galaxies in phase-space avoid the low-velocity region of phase-space, preferring the intense ram-pressure stripping areas. In other words, many of the jellyfish galaxies that sit outside the high-intensity ram-pressure stripping zone in Figure~\ref{PPS_jellys_byclM} are possibly being affected by substructures within the cluster and clumpy ICM. As mentioned earlier, we plan to present an in-depth analysis of the presence and effect of cluster substructure in the formation of jellyfish galaxies in a follow-up study.
Lastly, it is also possible that some of these galaxies, although have clearly been disturbed by ram-pressure, are not well represented by our models, or have not lost a significant fraction of gas ($>10\%$) to the ICM yet. %

If we group all stripping cases together (``Extreme stripping'', ``Stripping'' and ``Post-stripping''), $\sim 60$\% are inside or very near the plotted ram-pressure stripping ``cones''. The lowest coincidence occurs for low-mass galaxies in low-mass clusters. Finally, many of the galaxies classified as ``Unknown'' are also located inside or near this region, 
which suggests that ram-pressure could be at least one of the mechanisms responsible for their disturbed appearance.

Overall, the properties of the jellyfish galaxies, as revealed from MUSE data, together with their location in position vs. velocity phase-space conclusively show that ram-pressure stripping by the dense ICM is the main mechanism responsible for the jellyfish-like features in these galaxies. 
Moreover, our results support a scenario in which ram-pressure incrementally strips the gas from galaxies,  
with peak-stripping occurring as they approach pericentre.  Our findings are in agreement with the HI-stripping sequence presented in \citet{Yoon2017}.

\begin{figure}
\centering
\includegraphics[scale=0.48]{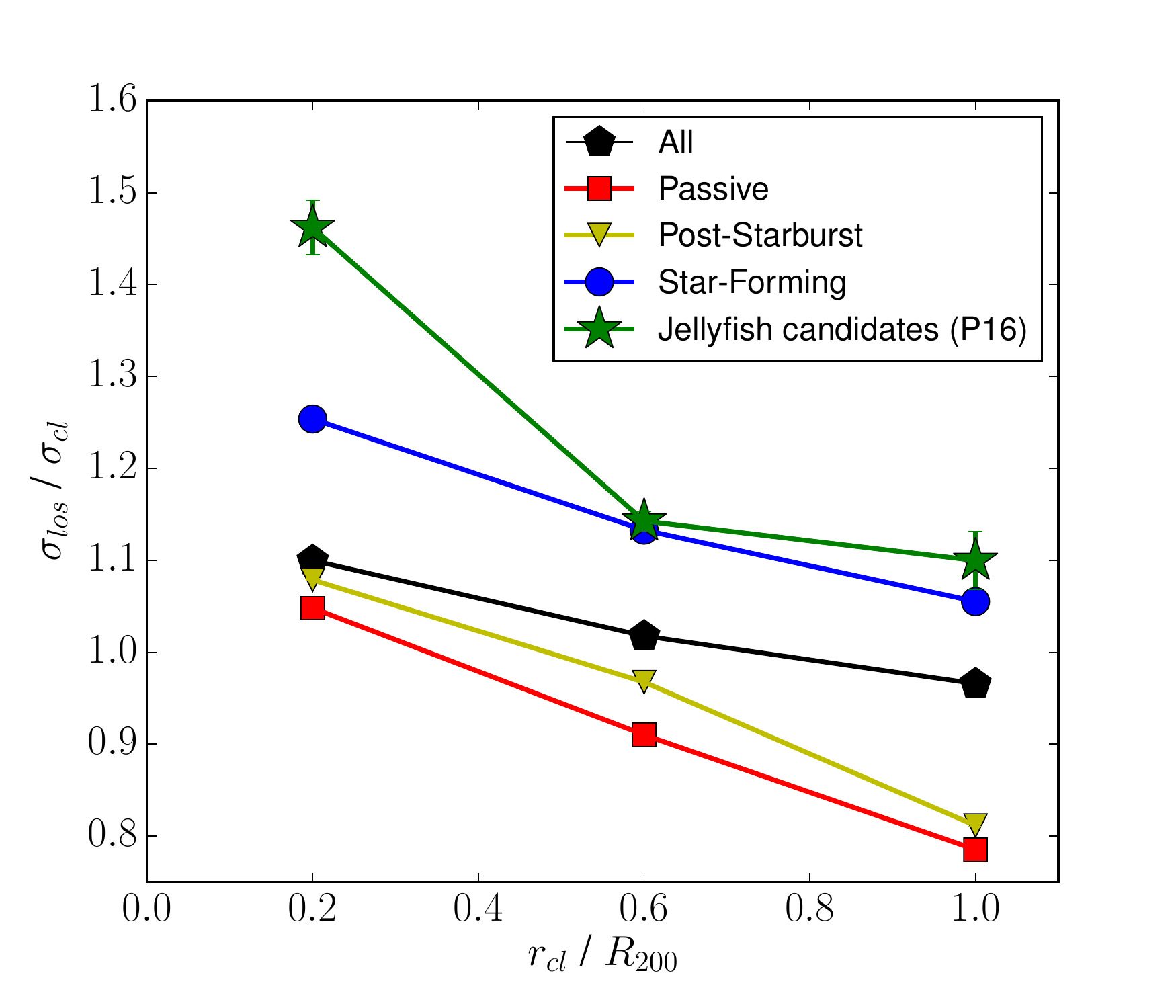}
\caption{Line-of-sight velocity dispersion radial profiles of all the cluster jellyfish candidates from P16, 
and the samples of passive, star-forming, and post-starburst galaxies from \citet{Paccagnella2017}, normalized by the velocity dispersion of the cluster as a function of distance from the cluster centre in units of $R_{200}$. The parent sample of WINGS/OmegaWINGS cluster galaxies is plotted in black for reference. We considered all candidate jellyfish (which include the confirmed cases from GASP) and did not differentiate by class to increase the number of galaxies per radial bin. 
Errors are jackknife standard deviations. For this plot we restricted the galaxy samples to the clusters with spectroscopic completeness $> 50\%$ and clustercentric distances $r_{cl} < R_{200}$ to minimize biases.   
\label{sigma}}
\end{figure}

\subsection{Cluster velocity dispersion}
\label{subsec:veld}

Additional information about the orbits of cluster galaxies may also be retrieved from the cluster velocity dispersion of different galaxy populations as a function of projected clustercentric distance \citep[see e.g][]{BivianoKatgert2004,Haines2015}. 
In Figure~\ref{sigma} the cluster velocity dispersion of the jellyfish galaxies (green stars) is compared with other galaxy populations defined in the following. 
We use here the spectral classification of WINGS/OmegaWINGS galaxies  presented in \citet{Paccagnella2017} who defined ``star-forming'' galaxies as those with H$\alpha$ or other emission lines in their spectra, and separated the galaxies without emission into ``Passive'' and ``Post-starburst''. The ``Passive'' type has weak H$\delta$ in absorption (H$\delta<3 \AA$, typical of K-type stars), while the ``Post-starburst'' displays a combination of signatures typical of both K and A-type stars with strong H$\delta$ in absorption ($H_{\delta}>3$ A).

In Figure~\ref{sigma}, the red line with the lowest cluster velocity dispersion at all clustercentric radii corresponds to the passive galaxies, and mostly traces the oldest population of cluster galaxies. The low velocity dispersion of these galaxies indicates they have ``virialized'' or settled into the cluster potential. 
The blue line corresponds to star-forming galaxies, that typically trace galaxies that have not been in the cluster for more than one or two pericentric passages, which explains the higher cluster velocity dispersion, typical of non-virialized cluster members. 
Now, if we separate the jellyfish galaxy candidates (of all classes) from the rest of the parent star-forming population we get the green line, which shows a dramatic increase of cluster velocity dispersion at low radii. In fact, the KS probability that the phase-space distribution of jellyfish candidates is drawn from the same distribution as star-forming galaxies (at $r_{cl} < R_{200}$) is very low (0.001\%). 
Assuming that the star-forming galaxies are the progenitors of jellyfish, the steep radial cluster velocity dispersion profile of jellyfish can be explained by highly radial orbits, and is consistent with the required conditions for ram-pressure stripping, which naturally occur near first pericentre. Moreover, the velocity distribution of our jellyfish galaxies is consistent with the steep cluster velocity dispersion profile of HI-deficient galaxies \citep{Solanes2001}. Our results are also consistent with the work by \citet{Vulcani2017}, who studied the orbits of  potentially ram-pressure stripped galaxies in intermediate-redshift clusters, selected to have offsets in the the peak of $H\alpha$ emission with respect to UV continuum. Using cosmological simulations as referece \citep[][]{Jiang2015}, they concluded that stripped galaxies are on first infall, and consist of the 25\% most radial orbits.

It is interesting to include the post-starburst galaxies in the comparison. Figure~\ref{sigma} shows that their cluster velocity dispersion profile (yellow line) is slightly above the passive galaxies. This was already seen in \citet{Paccagnella2017}, who 
concluded that the post-starburst galaxy population arises from a fast shut off of the star formation that changes the spectral features of the galaxy before it changes its color and morphology \citep[see also][]{Poggianti2004, Muzzin2014}. Considering the fast-acting effect of ram-pressure stripping it is likely that  post-starburst galaxies experienced significant gas stripping prior to obtaining post-starburst features. We will further investigate this idea in forthcoming papers by analyzing the post-starburst galaxies in the GASP sample.

\section{Summary and conclusions}
\label{sec:conclu}

In this paper we have presented the first large study of the orbital histories of jellyfish galaxies in clusters, utilizing the vast spectroscopic sample of cluster galaxies from the WINGS and OmegaWINGS surveys, the associated catalog of jellyfish candidates from P16, and the subset of confirmed jellyfish galaxies observed with MUSE by the ongoing GASP survey \citep{Poggianti2017a}. We list our main findings below. 
\begin{enumerate}[1)]
 \item We first characterize the sample of confirmed and candidate jellyfish galaxies and the control sample of non-stripped galaxies, which display a broad range of stellar mass and cluster mass. Jellyfish galaxies observed by MUSE were classified according to  their $H_\alpha$ morphology. The classification yielded four broad categories related to their apparent gas stripping stage: ``Stripping'' (one-sided asymmetries of the gas), ``Extreme-stripping'' (tails of stripped gas longer than the stellar body of the galaxy), ``Post-stripping'' (truncated $H_\alpha$ profiles relative to the stellar disk) and ``?'' (unclear asymmetry).  
 \item Using the parent sample of WINGS/OmegaWINGS clusters as reference, we constructed an analytic model of the ICM in two extreme cases: a low-mass cluster (Virgo, Cluster A) and a high mass cluster (A85, Cluster B). Similarly, we took an average low mass ($10^{9.6}M_{\odot}$, Galaxy I) and a high mass galaxy ($10^{10.7}M_{\odot}$, Galaxy II) from our sample to build two realistic models of the radial profile of the anchoring force in each case. By comparing the galaxy models with the cluster models it is possible to study the effect of ram-pressure in different scenarios. In  particular, we provide the \textit{recipe for tracing ram-pressure intensity in position vs. velocity phase-space diagrams of galaxies in clusters, and estimate the amount of total mass stripped from infalling galaxies.}
 \item We then utilize phase-space diagrams as diagnostic tool to infer the orbital histories of the cluster galaxies and the intensity of ram-pressure stripping exerted by the ICM. We find that:
 \begin{itemize}
 \item The distribution of jellyfish galaxy candidates in phase-space spans a broad range of projected clustercentric distances, but is preferentially shifted towards higher absolute line-of-sight velocities than the overall population of cluster galaxies at all clustercentric radii. This suggests that \textit{jellyfish galaxies were recently accreted into the clusters}. 
 \item When separating the confirmed jellyfish galaxies observed by GASP we find that the distribution of  ``Stripping'', ``Extreme-stripping'' and ``Post-stripping'' galaxies is consistent with \textit{incremental outside-in gas stripping of infalling galaxies}. The vicinity of the most extreme jellyfish to the cluster centre suggests that they are undergoing peak stripping near pericentre. Interestingly all the ``Extreme stripping'' galaxies have stellar masses $>10^{10}M_{\odot}$.
  \item  Finally, jellyfish galaxies show a steeper cluster velocity dispersion (especially at low projected clustercentric distances)
 than  passive and even star-forming galaxies, which indicates that  \textit{they are preferentially on radial orbits}. 
\end{itemize}
\end{enumerate}

Overall, our results reveal that jellyfish galaxies are a population of recently accreted cluster galaxies on highly radial orbits, that are stripped via ram-pressure as they pass through the dense cluster core at very high speeds. Since we detect significant gas stripping on first infall, we conclude that the theoretical prediction that gas removal is a fast ($\sim 1-2$ Gyr; first pericentric passage) and incremental outside-in effect driven by the ICM.  This is consistent with the stripping timescale found in cosmological simulations of cluster galaxies \citep[$\gtrsim 1$Gy;][]{Tonnesen2007} and the free-fall time of a galaxy moving from $1.5 \times R_{200}$  to $0.25 \times R_{200}$ ($1.1$~ Gry). The most spectacular jellyfish galaxies in our sample are high-mass galaxies located close to the cluster core (in projection). Simulations of ram-pressure stripping suggests that the observed long tails of ionized stripped gas are only possible to form under the most intense ram-pressure conditions. It is thus possible that only the massive galaxies were able to still have gas by the time they reached the vicinity of the cluster core and thus are the only ones displaying long bright tails.

The characterization of ram-pressure stripping in phase-space based on the cluster and galaxy models presented in Section~\ref{sec:rps_model} can serve as reference for other ram-pressure stripping studies (including published and future GASP papers). Although these models are simplistic and idealized, they have proven to be very good at predicting the amount of stripping for disk galaxies infalling into relatively regular galaxy clusters. In fact, the results presented in this paper are supported by cosmological \citep{Oman2013,Jaffe2015,Haines2015,Rhee2017} and hydrodynamical simulations \citep[e.g.][]{Roediger2006,TonnesenBryan2009,Steinhauser2016}, as well as similar studies of the distribution of HI-deficient and non-deficient galaxies in nearby clusters \citep{Solanes2001,Yoon2017}.
The known uncertainties in our ram-pressure stripping modeling introduced by the use of projected phase-space diagrams can be overcome when working with statistically large samples of galaxies.

Finally, it is important to note that the results presented in this paper are based on a sample of mostly relaxed clusters. 
There is evidence however that in dynamically disturbed clusters extreme ram-pressure stripping events can be abundant \citep{Vijayaraghavan2013,Jaffe2016,McPartland2016}. It is reasonable to conclude that in a hierarchical Universe, the conditions where extreme ram-pressure stripping occurs not only depend on the  properties of the galaxy, mass of the host cluster, and orbital parameters (as described in this paper), but also on the dynamical state of the host cluster/group. We plan to present a dedicated study of the effect of cluster sub-structuring in the formation of jellyfish galaxies in a future publication.  

\section*{Acknowledgments}

Based on observations collected at the European Organisation for Astronomical Research in the Southern Hemisphere under ESO programme 196.B-0578. We thank Jinsu Rhee for providing the simulations data used in this paper and Graeme Candlish for useful discussions. 
We acknowledge financial support from PRIN-SKA 2017 (PI Hunt). 
B.V. acknowledges the support from an Australian Research Council Discovery Early Career Researcher Award (PD0028506).

\bibliographystyle{mn2e}

\end{document}

%% file: mycommands.tex
\def\HI{\ifmmode{\rm HI}\else{H\/{\sc i}}\fi}

\def\lsun{\ifmmode{{\mathrm L}_{\odot}}\else{L$_{\odot}$}\fi}

\def\msun{\ifmmode{{\mathrm M}_{\odot}}\else{M$_{\odot}$}\fi} 
\def\msunpc2{\ifmmode{{\mathrm M}_{\odot} \, {\mathrm{pc}}^{-2}}\else{M$_{\odot} \, {\mathrm {pc}}^{-2}$}\fi}

\def\kms{\ifmmode{{\mathrm{km \, s^{-1}}}}\else{${\mathrm{km \, s^{-1}}}$}\fi}

\def\la{\mathrel{\mathchoice {\vcenter{\offinterlineskip\halign{\hfil
$\displaystyle##$\hfil\cr<\cr\sim\cr}}}
{\vcenter{\offinterlineskip\halign{\hfil$\textstyle##$\hfil\cr
<\cr\sim\cr}}}
{\vcenter{\offinterlineskip\halign{\hfil$\scriptstyle##$\hfil\cr
<\cr\sim\cr}}}
{\vcenter{\offinterlineskip\halign{\hfil$\scriptscriptstyle##$\hfil\cr
<\cr\sim\cr}}}}}
\def\ga{\mathrel{\mathchoice {\vcenter{\offinterlineskip\halign{\hfil
$\displaystyle##$\hfil\cr>\cr\sim\cr}}}
{\vcenter{\offinterlineskip\halign{\hfil$\textstyle##$\hfil\cr
>\cr\sim\cr}}}
{\vcenter{\offinterlineskip\halign{\hfil$\scriptstyle##$\hfil\cr
>\cr\sim\cr}}}
{\vcenter{\offinterlineskip\halign{\hfil$\scriptscriptstyle##$\hfil\cr
>\cr\sim\cr}}}}}

\def\aj{AJ}
\def\araa{ARA\&A}
\def\apj{ApJ}
\def\apjl{ApJ}
\def\apjs{ApJS}
\def\aap{A\&A}
\def\aapr{A\&A~Rev.}
\def\mnras{MNRAS}
\def\pasp{PASP}
\def\nat{Nature}